\pdfoutput=1
\documentclass[preprint]{elsarticle}

\usepackage[utf8]{inputenc}
\usepackage[T1]{fontenc}
\usepackage[english]{babel}
\usepackage{textcomp}

\usepackage[hyphens]{url}
\biboptions{sort&compress, square, comma}
\usepackage[breaklinks=true, linkcolor=blue, citecolor=blue, colorlinks=true]{hyperref}

\usepackage{latexsym, amsmath, amssymb}
\usepackage{graphicx, color, xcolor, wrapfig, subcaption}
\usepackage{listings}

\usepackage{nicefrac}

\usepackage[binary-units]{siunitx}

\definecolor{dkgreen}{rgb}{0,0.6,0}
\definecolor{gray}{rgb}{0.5,0.5,0.5}
\definecolor{mauve}{rgb}{0.58,0,0.82}

\def\CC{{C\nolinebreak[4]\hspace{-.05em}\raisebox{.4ex}{\footnotesize ++}}}

\usepackage[textsize=footnotesize]{todonotes}
\graphicspath{ {images/} }

\newcommand{\TheTitle}{Accelerating solutions of one-dimensional unsteady PDEs with GPU-based swept time-space decomposition}

\begin{document}

\begin{frontmatter}

\title{\TheTitle}

\author[OSU]{Daniel J.~Magee}

\author[OSU]{Kyle E.~Niemeyer\corref{cor1}}
\ead{kyle.niemeyer@oregonstate.edu}

\cortext[cor1]{Corresponding author}
\address[OSU]{School of Mechanical, Industrial, and Manufacturing Engineering, Oregon State University, Corvallis, OR, USA}

\begin{abstract}
The expedient design of precision components in aerospace and other high-tech industries
requires simulations of physical phenomena often described by partial differential
equations (PDEs) without exact solutions.
Modern design problems require simulations with a level of resolution difficult
to achieve in reasonable amounts of time---even in effectively parallelized solvers.
Though the scale of the problem relative to available computing power is the greatest
impediment to accelerating these applications, significant performance gains can be
achieved through careful attention to the details of memory communication and access.
The swept time-space decomposition rule reduces communication between sub-domains
by exhausting the domain of influence before communicating boundary values.
Here we present a GPU implementation of the swept rule, which modifies the algorithm
for improved performance on this processing architecture by prioritizing use of private
(shared) memory, avoiding interblock communication, and overwriting unnecessary values.
It shows significant improvement in the execution time of finite-difference solvers for
one-dimensional unsteady PDEs, producing speedups of
\SIrange{2}{9}{$\times$} for a range of problem sizes, respectively, compared with
simple GPU versions and \SIrange{7}{300}{$\times$} compared with parallel CPU versions.
However, for a more sophisticated one-dimensional system of equations discretized with a
second-order finite-volume scheme, the swept rule performs \SIrange{1.2}{1.9}{$\times$}
worse than a standard implementation for all problem sizes.
\end{abstract}


\end{frontmatter}


\section{Introduction}

High-fidelity computational fluid dynamics (CFD) simulations are essential for developing
aerospace technologies such as rocket launch vehicles and jet engines.
This project aims to accelerate such simulations to approach real-time execution---simulation
at the speed of nature---in accordance with the high-performance computing development goals
set out in the CFD Vision 2030 report~\cite{slotnick:2014}.
Classic approaches to domain decomposition for parallelized, explicit, time-stepping partial
differential equation (PDE) solutions incur substantial computational performance costs
from the communication between nodes required every timestep.
This communication cost consists of two parts: latency and bandwidth, where latency
is the fixed cost of each communication event and bandwidth is the variable cost
that depends on the amount of data transferred.
Latency in inter-node communication is a fundamental barrier to this goal, and advancements
to network latency have historically been slower than improvements in other computing
performance barriers such as bandwidth and computational power~\cite{PattersonLatency}.
Performance may be improved by avoiding external node communication until
exhausting the domain of dependence, allowing the calculation to advance multiple timesteps
while requiring a smaller number of communication events.
This idea is the basis of swept time-space decomposition~\cite{alhubail:16jcp,Alhubail:2016arxiv}.

Extreme-scale computing clusters have recently been used to solve the compressible
Navier--Stokes equations on over 1.97 million CPU cores~\cite{BM_BigSolution}.
The monetary cost, power consumption, and size of such a cluster impedes the realization
of widespread peta- and exa-scale computing required for real-time, high-fidelity, CFD simulations.
While these are significant challenges, they also provide an opportunity to develop
new tools that increase the use of the available hardware resources.
As the authors of CFD Vision 2030 note, ``High Performance Computing (HPC) hardware
is progressing rapidly and is on the cusp of a paradigm shift in technology that may
require a rethinking of current CFD algorithms and software''~\cite{slotnick:2014}.
Using graphics processing unit (GPU) hardware as the primary computation device or
accelerator in a heterogeneous system is a viable, expedient option for high-performance
cluster design that helps mitigate these problems.
For this reason, GPUs and other emerging coprocessor architectures are increasingly
used to accelerate CFD simulations~\cite{Niemeyer:2014hn}.

GPU technology has improved rapidly in recent years; in particular, NVIDIA GPUs have progressed from Kepler to Pascal architecture in four years.
This development doubled and tripled peak single- and double-precision performance, respectively~\cite{PascalWP:2016}.
In addition, the presence of GPUs in clusters, such as ONRL's Titan supercomputer, has become increasingly common in the last decade.
These advances have driven the development of software capable of efficiently using and unifying the disparate architectures~\cite{Witherden20143028}.
Although the ultimate motivation of our work is accelerating the solution of PDEs---particularly
relevant to fluid flow---on distributed-memory systems, in this work we focused on a single GPU-accelerated
node\slash workstation in the design of the algorithms and associated software.
By investigating the effectiveness of the swept rule on a workstation, we provide results
that can be applied to simulations on a single machine as well as an initial framework for understanding
the performance of the swept rule on heterogeneous computing systems.

The swept rule operates on a simple principle: do the most work possible
on the values closest to the processor before communicating.
In practice, this directive results in an algorithm that advances the solution in
time at all spatial points using locally accessible stencil values at the previous timestep.
Because the data closest to the processor is also the least widely accessible, the
strict application of this principle does not always provide the best performance,
but it is a useful heuristic for implementing the procedure and analyzing its performance.

This study presents an investigation of the performance characteristics of three
swept rule implementations for a single GPU in a workstation.
These procedures are tested on three one-dimensional PDEs with numerical schemes
of varying complexity and compared with the performance of parallel CPU algorithms
and unsophisticated GPU versions. The (next) Section~\ref{sec:Rwork} describes recent
work on partitioning schemes for PDEs and communication-avoiding algorithms,
especially as applied to GPUs. Section~\ref{sec:GPUArch} gives a brief overview of
the GPU architecture, particularly the thread and memory hierarchies.
Section~\ref{sec:methods} discusses the swept rule, and our adjustments to the original
algorithm in response to the details of GPU architecture.
Section~\ref{sec:schemes} describes the swept rule implementation in detail.
In Section~\ref{sec:results} we present the results of the tests and, lastly, draw further conclusions in Section~\ref{sec:conclude}.

\section{Related work}\label{sec:Rwork}

Alhubail and Wang introduced the swept rule for explicit, time-stepping, numerical schemes applied to PDEs~\cite{MaithamRepo,alhubail:16jcp,Alhubail:2016arxiv}, and our work takes their results and ideas as its starting point.
The swept rule is closely related to cache optimization techniques, in particular those that use geometry to organize stencil update computation such as parallelograms~\cite{Strzodka} and diamonds~\cite{MalasHager}.
The diamond tiling method presented by Malas et al.~\cite{MalasHager} is similar to the swept rule but uses the data dependency of the grid to improve cache usage rather than avoid communication.
Concepts such as stencil optimization using domain decomposition on various architectures that are fundamental to this study are explored by Datta et al.~\cite{VolkovDatta2008}. Their work explores comparisons between parallel GPU and CPU architectures and tunes the stencil algorithm with nested domain decomposition.
The swept rule also has elements in common with parallel-in-time and communication-avoiding algorithms.

Parallel-in-time methods~\cite{Gander2015}, such as multigrid-reduction-in-time (MGRIT)
algorithms~\cite{falgout2014parallel}, accelerate PDE solutions with time integrators that
overcome the interdependence of solutions in the time domain, allowing parallelization
of the entire space-time grid. These methods calculate the solution over the space-time
domain using a coarse grid and iterate over successively finer grids to achieve the desired accuracy.
The use of coarse grids in parallel-in-time methods reduces efficiency and accuracy
when applied to nonlinear systems~\cite{alhubail:16jcp}. This shortcoming is intuitive: since
chaotic, nonlinear systems may suddenly change in time, and coarse grids are prone to aliasing,
the required grid granularity diminishes gains in performance.
The swept rule arises from the same motivation,
but does not seek to parallelize the computation in time or vary dimensions during the process.

The swept rule does not alter the numerical scheme; it decomposes the domain and
organizes computation.
That is, compared to a classic domain decomposition, the swept rule performs the same
operations in a different order and location.
In this way communication-avoiding algorithms share many implementation details with the swept rule. Recent developments in communication-avoiding algorithms for GPUs have generally focused on applications involving matrices such as QR and LU factorization.
The LU factorization algorithm presented by Baboulin et al.~\cite{BABOULIN201217} is motivated by the increasing use of GPU accelerators in large-scale, heterogeneous clusters.
This method splits tasks between the GPU and CPU, minimizing communication between devices.
This allows the communication and computation performed on each device to overlap, so all
data transfer occurs asynchronously with computation. We explore this approach---overlapping
data transfer with hybrid computation---in this article.
The motivation and structure of our study is comparable to the work of Anderson et al~\cite{Anderson}.
They developed methods for arranging and tuning computation for single general-purpose GPU (GPGPU) in a desktop workstation, without altering the basic QR factorization algorithm.
Similarly, this study focuses on adapting the swept rule to a single GPU.
The swept rule is a strategy for arranging the computational path of explicit numerical methods, and this work seeks to design the data structures and operations used in that path to achieve the best performance on GPUs.

\section{GPU architecture and memory} \label{sec:GPUArch}

The impressive parallel processing capabilities of modern GPUs resulted from
architecture originally designed to improve visual output from a computer.
GPU manufacturers and third-party authors~\cite{Owens:2008ku,cudaProgGuide,Brodtkorb:2013hn,EngineerCuda}
have described the features of this architecture in great detail, while others
discussed general best practices for developing efficient GPU-based algorithms~\cite{Cruz:2011gc,Niemeyer:2014hn}.
Particular aspects of this architecture, such as the unique and accessible memory hierarchy,
are at the core of this work, so some explanation of its relevant elements is necessary
before describing the details of the implementation.

Programs that run on the GPU can be implemented using several software packages, the most
common of which are the OpenCL and OpenACC frameworks, and the CUDA parallel computing platform.
These packages use different nomenclatures and are compatible with different hardware types.
In this project all programs use CUDA, which is exclusively compatible with NVIDIA GPUs;
therefore, all descriptions of GPU hardware presented here use the CUDA nomenclature.
CUDA programs consist of functions, referred to as kernels, launched from a C\slash\CC{} host program.
The CPU executes the host code and specifies the size and number of blocks when calling a
kernel, the stream (queue) in which the kernel will be launched, and the amount of
shared memory to be allocated per block at runtime. All threads in a warp---a group of
\num{32} threads that execute as a single-instruction multiple thread (SIMT) unit---must
be in the same block, so for good practice blocks should launch with some multiple of 32 threads.

The information presented here is valid for all NVIDIA GPUs with compute capability \num{3.0}
or higher (i.e., Kepler architecture or later). The device used in this study is a
Tesla K40c GPGPU, compute capability 3.5.
This device contains 15 streaming multiprocessors, each capable of processing 64 warps
of 32 threads, or 2048 total threads, at once~\cite{PascalWP:2016}.
A maximum of \num{16} blocks may concurrently reside on a streaming multiprocessor;
blocks may not be split between streaming multiprocessors.
While each streaming multiprocessor can support 2048 resident threads, in practice their
capacity is often lower because each thread or block makes demands on limited memory
resources---most notably the shared memory and registers.
Each streaming multiprocessor on the Tesla K40c has \SI{48}{\kilo\byte} of shared memory
and 65536 registers available. Registers offer the fastest access, but are the most limited memory type
and are private to each thread, but can be accessed by other threads in the same warp
using shuffle operations available on devices with compute capability 3.5 or higher.
Shared memory is a controllable portion of the L1 cache accessible only to threads within a block.
As a result, for a thread to read a value stored in shared memory in a different block,
a thread with access to that value must write the value to global memory where
the reader thread has access. Global memory is the slowest and most plentiful memory
type, and where data copied from the host program resides.
Global memory stores all variables passed to a kernel and large arrays declared therein.

Other memory types in the CUDA memory hierarchy include constant, texture, and surface;
of these, the work presented here only uses constant memory. Constant memory is read-only,
available to all kernels for the lifetime of an application, and quick to access
when all threads access the same location. This makes it a convenient and performance conscious
choice for storing constant values of the governing equations calculated at runtime~\cite{cudaProgGuide}.

\section{Methodology}\label{sec:methods}

\subsection{Experimental method}

The primary goal of this study is to compare the performance of the swept rule
to a simple domain decomposition scheme, referred to as \texttt{Classic}, on a GPU.
A domain decomposition scheme is a way of splitting up a large problem so tasks can
be shared by discrete workers working on distinct parts of the problem.
In this case the decomposition scheme divides the space-time domain of the PDE solution.
We will compare the performance of these methods by the time cost of each timestep,
including the time required to transfer data to\slash from the GPU.
While encoding the \texttt{Classic} is relatively straightforward, finding the best
approach for the swept rule on the GPU presents a more subtle problem.

In the original swept rule approach~\cite{alhubail:16jcp}, the spatial domain
is partitioned into independent pieces called ``nodes'' that correspond to compute
nodes on a distributed system with private memory spaces.
A major concern in adapting this nodal analogy to a single GPU is the type of memory
allocated for the working array, the container for the values that are the solution
to and the basis for each timestep. Several available approaches exist to map the
original analogy for a node to a single GPU; here, we will explore three of them:
\texttt{Shared}, \texttt{Hybrid}, and \texttt{Register} (which we will describe in detail in Section~\ref{sec:schemes}).

In all approaches we map one thread to one spatial point.
Handling more than one spatial point per thread would allow for larger nodes but would require more resources and complicate the procedure without reducing thread idleness.
Additional swept rule properties could be adjusted for potential implementation variants such as the data structure to hold the working values, and the method to globally synchronize threads.
However, for the purposes of this study, we did not vary these attributes.
In all cases in this study the working array is a standard, one-dimensional C array with two flattened rows; and kernel calls are implicitly synchronized by returning control to the queue in the host program.

\subsection{First-order domain of dependence} \label{sec:OrderOne}

\begin{figure}[!b]
	\centering
	\begin{subfigure}[t]{.69\textwidth}
		\centering
		\includegraphics[width=\textwidth]{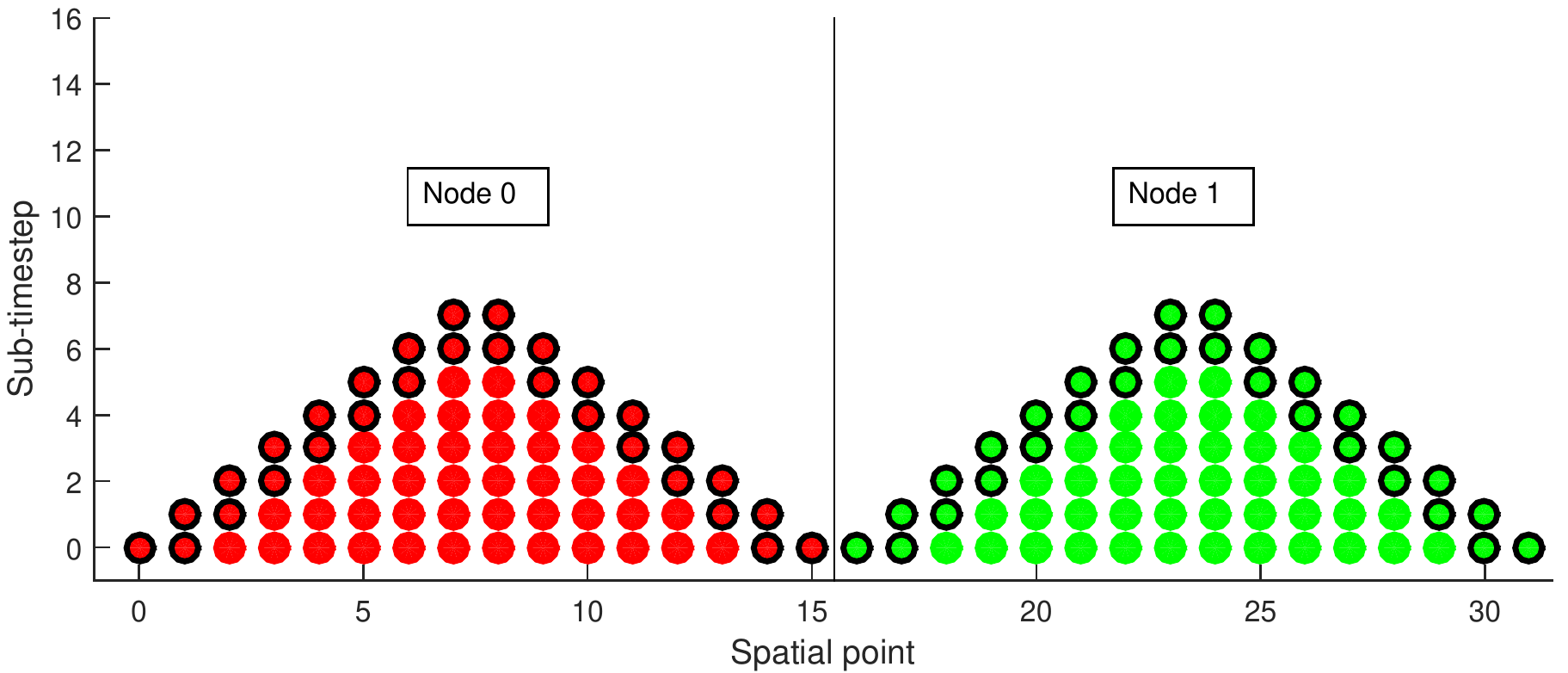}
		\caption{The first step of the swept rule.
		Values at $t=0$ are split between nodes \num{0} and \num{1}, which compute solutions
        in their domain of dependence, a triangle in the space-time plane.
		The edge values are collected in global arrays $L_0$/$R_0$ and $L_1$/$R_1$.}
		\label{f:firstorder1}
	\end{subfigure}
	\hfill
	\begin{subfigure}[t]{.69\textwidth}
		\centering
		\includegraphics[width=\textwidth]{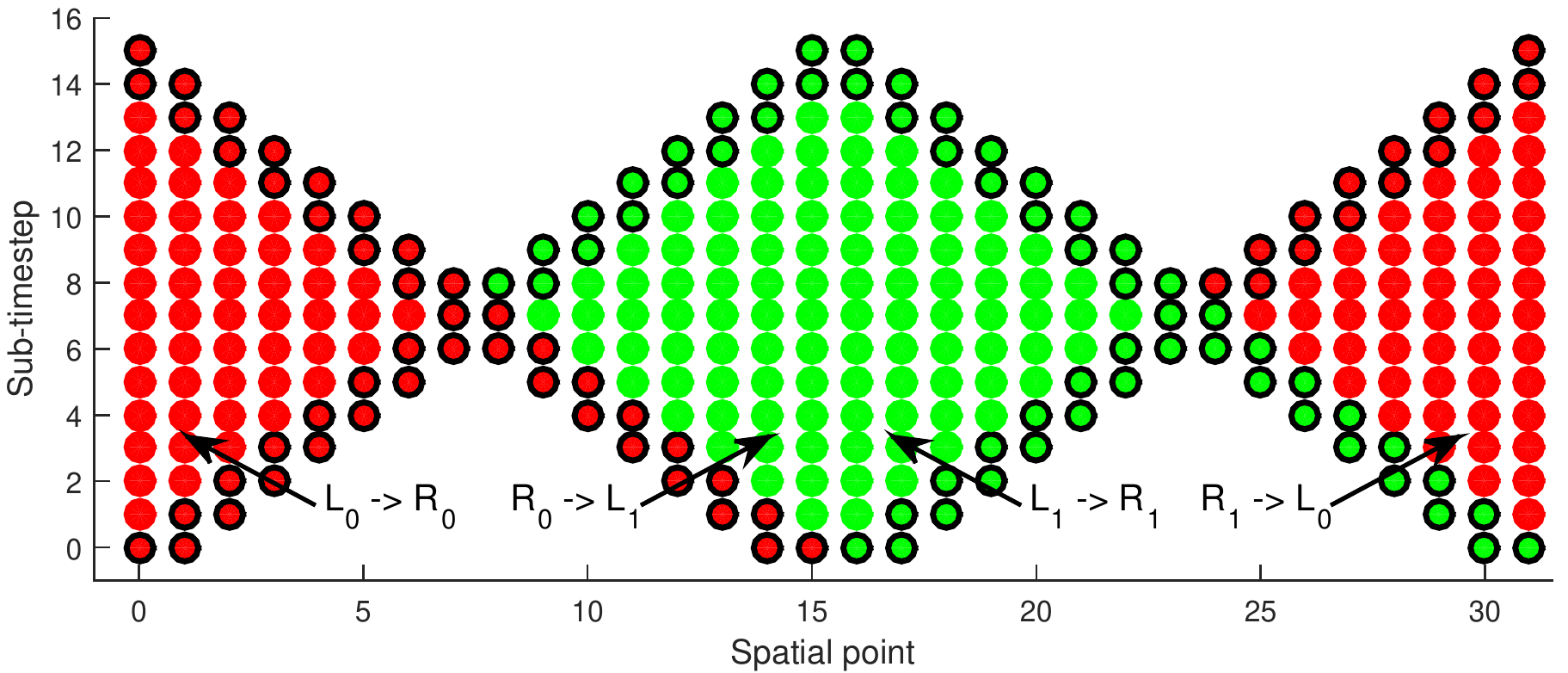}
		\caption{The second step in the swept rule.
		The nodes pass their right edge to the neighboring node.
		The passed values become the initial left edge, and the left edge from the previous stage becomes the right edge.
		Each node advances through their domain of dependence, a diamond in space-time.}
		\label{f:firstorder2}
	\end{subfigure}
	\caption{The first two steps of the swept rule for a numerical scheme with a first-order domain of dependence.
	$L_i/R_i$ refer to left/right arrays of node $i$, which collect edge values shown with thick bordered dots~\cite{FigsShared}.}
	\label{f:firstorder}
\end{figure}

A domain of dependence is a region on the space-time grid that can be completed by a numerical scheme with some set of initial values.
For the purposes of this project, numerical schemes consist of stencil operations, where a value at a
grid point is updated based on the weighted contribution of values at grid points in the vicinity.
A three-point stencil uses values at neighboring grid points only, and the timestep of any
numerical scheme can be decomposed into a series of sub-timesteps that only require a
three-point stencil~\cite{WangDecomp}.
The numerical scheme defines the order of the domain of dependence: it increases by one
for every two sub-timesteps required per timestep.
Therefore, a domain of dependence is \emph{first-order} if all sub-timesteps in the
numerical scheme use a three-point stencil, and intermediate values are required
no more than two steps after they are calculated.
The initial incarnation of the swept rule presented by Alhubail and
Wang~\cite{alhubail:16jcp,MaithamRepo} decomposes multi-step timesteps and large
stencils into sub-timesteps with a three-point stencil.
This regularizes the procedure and ensures that all equations and schemes can be
evaluated using a swept decomposition with a first-order domain of dependence,
but requires more memory to store the intermediate values that result from each sub-timestep.
In order to conserve limited private memory, for the GPU-based swept rule a first-order
domain of dependence is applicable to schemes that require two or fewer three-point stencil sub-timesteps per timestep.
Figure~\ref{f:firstorder} shows the first two stages of the swept rule using $k = 2$ nodes with $n = 16$ spatial points.

At the start of the first step, shown in Figure~\ref{f:firstorder1}, the initial
conditions are passed to the kernel and each node evaluates the solution at as many points as possible in the space-time grid.
The initial domain of dependence forms a triangle.
The working array stores the solutions as each node steps through time and is maintained in a fast memory space private to node member threads.
When each node cannot advance any further it's necessary for each to pass one edge to a neighboring node and retain the values of the other edge.
Figure~\ref{f:firstorder2} shows how the second step proceeds from the first using the edge values.

\begin{figure}[!b]
	\centering
	\begin{subfigure}[t]{.69\textwidth}
		\centering
		\includegraphics[width=\textwidth]{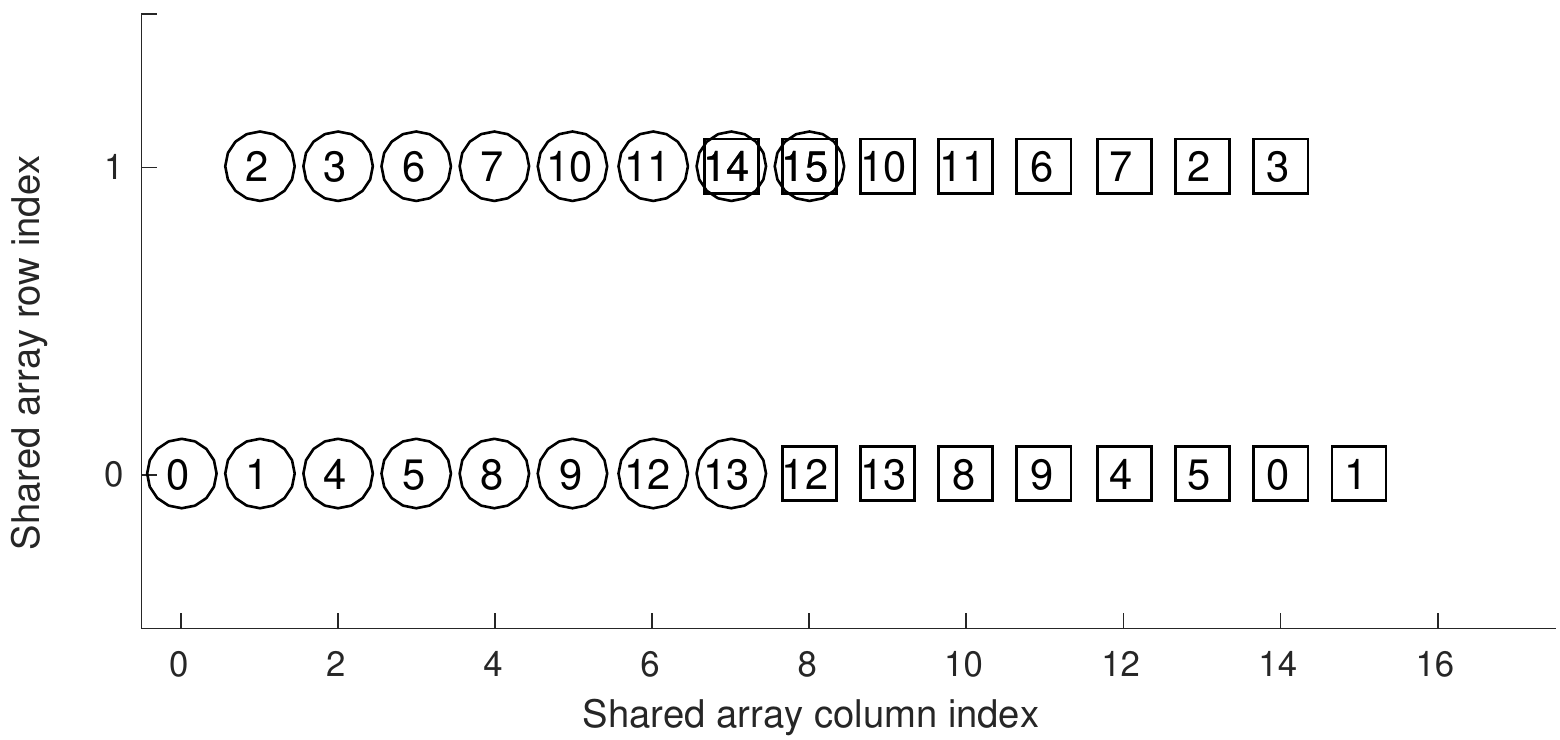}
        \caption{At the end of a kernel (completion of an inverted triangle), the working array is stored and passed from private memory to global memory.
        The location shows the index of the working array in (private) shared memory. The number in the shape refers to the
        offset global memory index where the working value is passed.
        }
		\label{f:extraction}
	\end{subfigure}
	\hfill
	\\
	\begin{subfigure}[b]{.69\textwidth}
		\includegraphics[width=\textwidth]{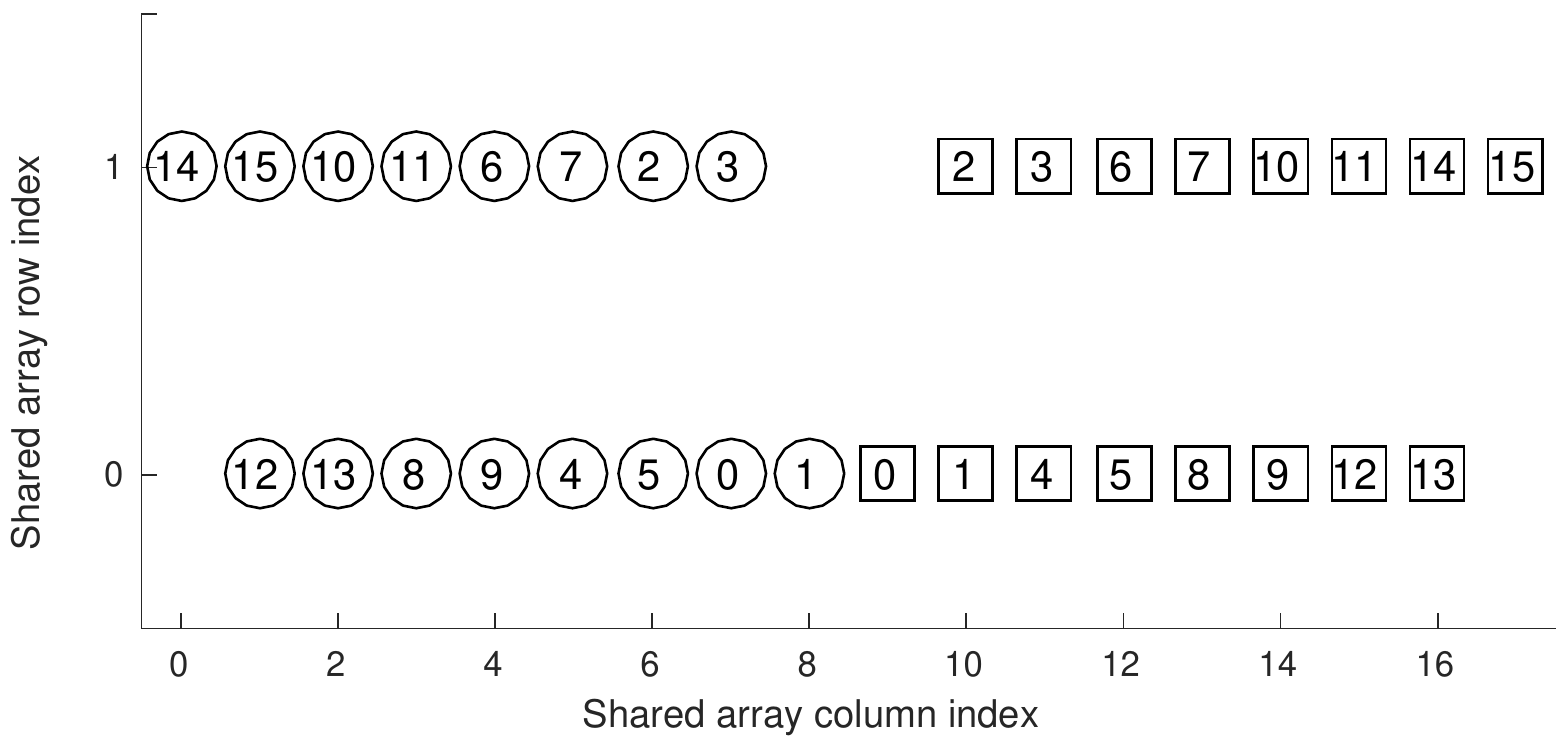}
		\caption{Edges are reinserted to seed the inverted triangle of a new swept cycle.
        The location shows the index of the working array in (private) shared memory receiving the global memory value denoted
        by the number.
        }
		\label{f:insertion}
	\end{subfigure}
	\caption{The procedure for edge passing shown in Figure~\ref{f:firstorder2}.
		The global arrays $L_i$ and $R_i$ are represented by circles and squares, respectively, and the numbers in those shapes represent the position of the value in the global array.
		The axes describe the location of the value in the working array~\cite{FigsShared}.}
	\label{f:collection}
\end{figure}

For the nodes to communicate as Figure~\ref{f:firstorder} illustrates, the values on the edges must be passed to global arrays available to all nodes since any sufficiently fast memory location is private to a subset of threads terminated on kernel exit.
Consequently, any data necessary to continue the computation must be shared
between nodes using an intermediary container.
The information needed to begin a new nodal cycle, including the values retained
by each node, must be read in from, and out to, global memory at the beginning and end of each kernel, respectively.
Figure~\ref{f:extraction} shows how the working array values are stored and passed to the global arrays.

Figure~\ref{f:firstorder2} illustrates the reason the two edges are stored
individually as $L_i$ and $R_i$: only one of the edges is passed between nodes.
After the first step, the right edge is passed to its right neighbor node; the left edge is stationary.
At the beginning of the subsequent kernel, $L_i$ and $R_i$ are swapped and reinserted
into the working array to seed the next progression with one data transfer event
from global to private memory as shown in Figure~\ref{f:insertion}.
When the right edge is passed between nodes, node $0$ is split across the spatial
boundary and must apply the boundary conditions at its center.

The swept rule allows the computation to advance by $n$ timesteps with two, rather
than $n$, global memory accesses, where $n$ is the number of threads per block (or
the number of spatial points per node). The procedure advances by passing values in the
alternating directions and zig-zagging the location of the nodes in this fashion until the simulation is complete.
Since the diamonds shown in Figure~\ref{f:firstorder2} do not store all the values
at a single timestep, the simulation can only output values when a complementary
triangle is computed and the final $n$-length local tier is returned.
This kernel can only be called after the values are passed to the left, so the
results can only be read out every $n$th timestep.

Although we already described the working array and showed in Figure~\ref{f:collection}
how the relevant values are communicated between the nodes, it is instructive to
outline the performance concerns that motivate this arrangement.
As Section~\ref{sec:GPUArch} describes, the number of resident threads on a streaming
multiprocessor depends on the GPU architecture and resources requested at kernel launch.
For instance, storing every double-precision value in the triangle shown in
Figure~\ref{f:firstorder1} in shared memory in a kernel with \num{512} threads per
block would require $8 \times 65792 = \SI{514}{\kilo\byte}$ of shared memory---this
is over 100 times greater than the \SI{48}{\kilo\byte} limit for NVIDIA GPUs with Kepler architecture.
The maximum number of threads per streaming multiprocessor would be limited to 128,
which would negatively impact program performance.

Figure~\ref{f:firstorder1} shows that the interior of the triangle is only
needed to progress to the next timestep, and that edges on even and odd tiers
do not overlap in the spatial domain.
Thus, the triangle may be stored as a matrix with two rows, where the first and
second rows contain the even and odd sub-timesteps results, respectively.
The interior values are overwritten once they are used, and only the edge values remain.
Figure~\ref{f:extraction} shows the result of a local computation with this method.
The last two values, the tips of each triangle, are copied into both arrays.

At the start of the next kernel, the left and right arrays are inserted into the
working array to seed successive calculations as Figure~\ref{f:insertion} shows.
The edges of the previous cycle's first row are moved to the center, and the center
of the current top row is now left open for the first tier of the inverted triangle.
Each row requires space for $n+2$ values because two edge values are required on
either side to complete the stencil for the longest row where $n$ values are computed.
The computation proceeds by filling the empty two indices on the top row, overwriting
the bottom row's middle four indices, and so on. In contrast to the memory demands of
storing the entire nodal computation, this method uses only $2\times(n+2)$ values.
For a block with \textit{n} = 512 threads, this requires $8 \times 1028 \approx \SI{8}{\kilo\byte}$.
In general, a stencil with width $k$ would require storing $2\times(n+k-1)$
values in the working array.
By reducing the amount of shared memory to only \SI{8}{\kilo\byte}, the kernel is
less limited and achieves higher occupancy, the number of threads each streaming
multiprocessor is capable of handling simultaneously.

\subsection{Higher-order domain of dependence} \label{sec:OrderTwo}

\begin{figure}[!bt]
	\centering
	\begin{subfigure}[t]{.7\textwidth}
		\centering
		\includegraphics[width=\textwidth]{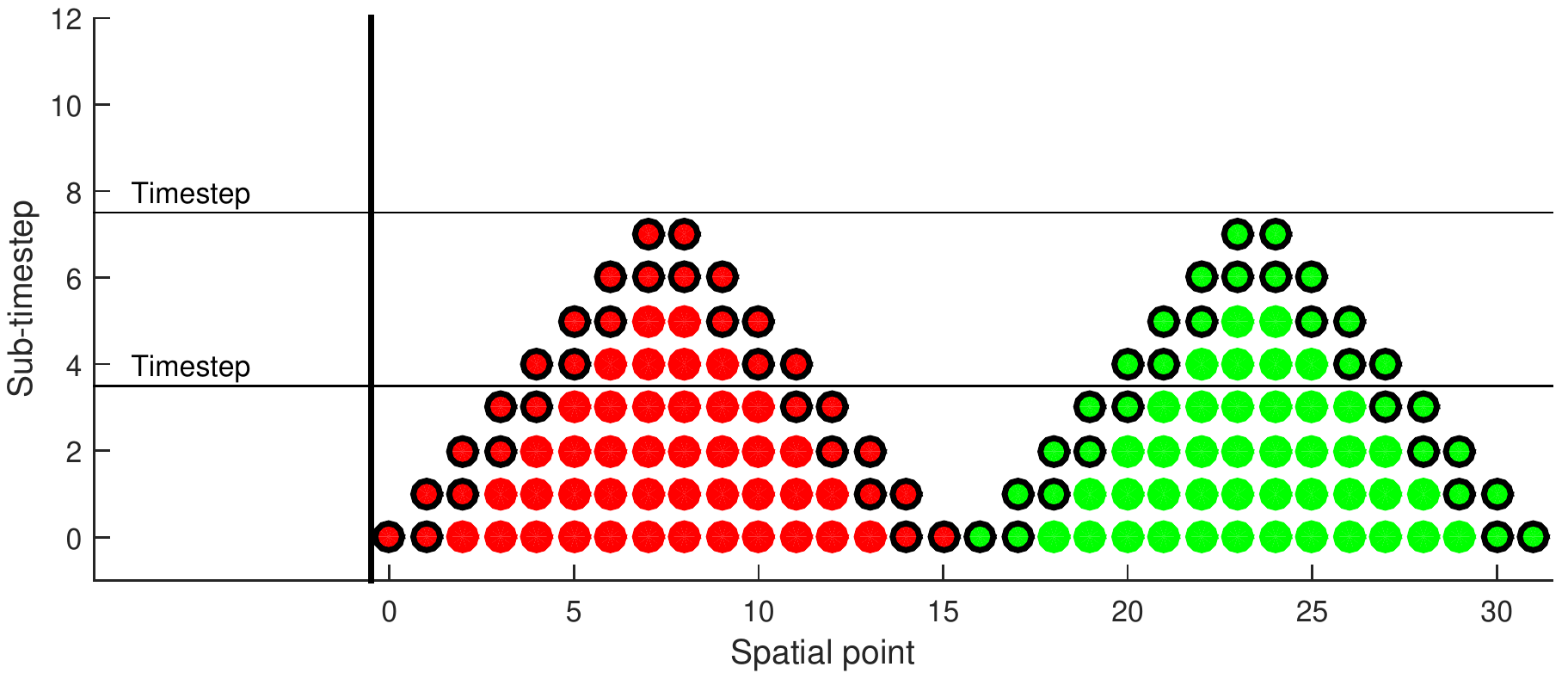}
		\caption{First step of swept rule for discretization with four, three-point stencil sub-timesteps per timestep.}
		\label{f:secondprob}
	\end{subfigure}
	\\
	\begin{subfigure}[b]{.7\textwidth}
		\includegraphics[width=\textwidth]{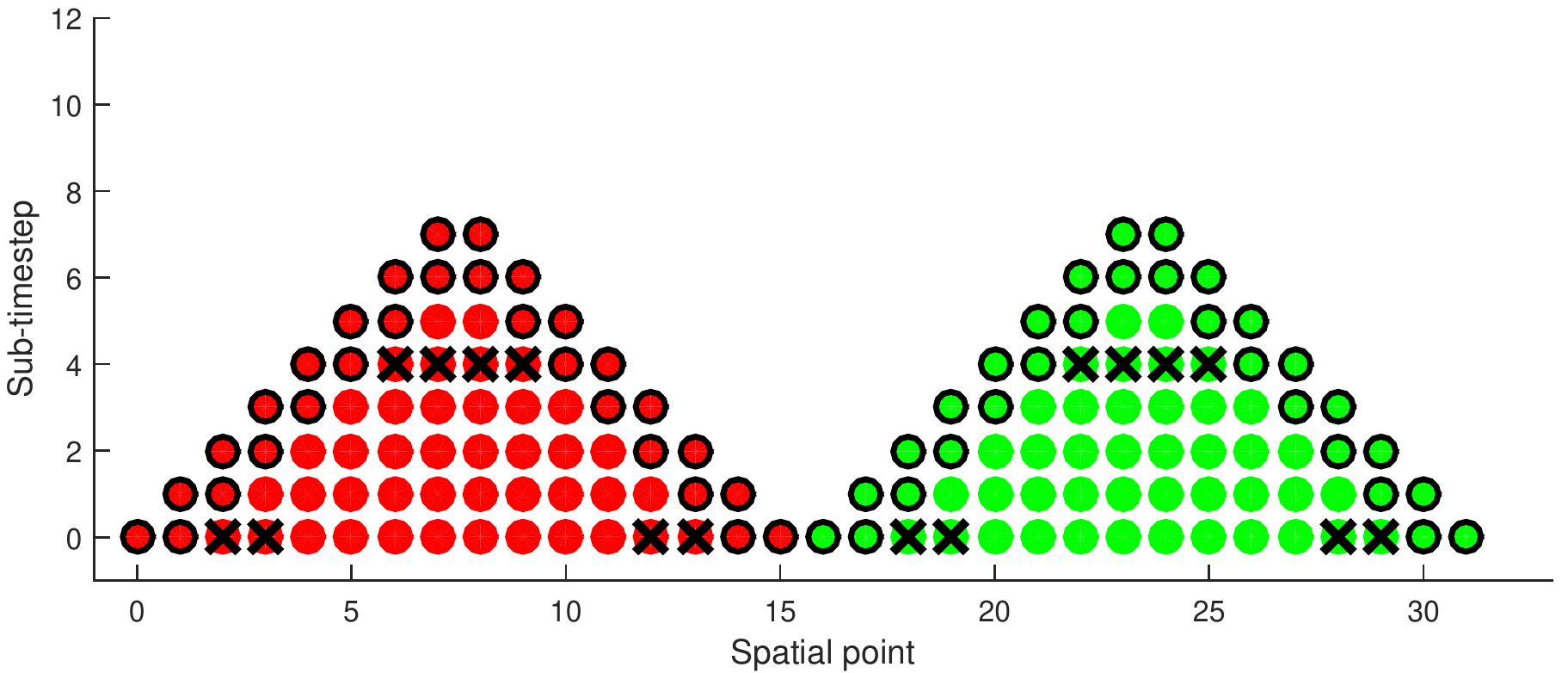}
		\caption{Using four sub-timesteps per timestep will overwrite values marked with an ``$\times$'' before they are needed.}
		\label{f:secondexes}
	\end{subfigure}
	\caption{Conflicts in the domain of dependence for discretizations requiring more than two sub-timesteps per timestep~\cite{FigsShared}.}
	\label{f:secondorderProblem}
\end{figure}

The first-order domain of dependence suffices for relatively simple problems.
However, more complicated problems with elements such as nonlinear equations,
discontinuities, or higher-order derivatives require more sophisticated procedures.
The original swept rule program calculates and stores intermediate values at sub-timesteps for higher-order schemes to avoid using larger stencils~\cite{alhubail:16jcp}.
Breaking a timestep into a series of sub-timesteps allows any numerical scheme for any equation of the same dimension to be decomposed in the same way since all stages in the computation depend on the minimum stencil, that is, only the neighbors of the current spatial point~\cite{WangDecomp}.
For example, a second-order in time midpoint method applied to a fourth-order differential equation would require four sub-timesteps per timestep.
The first sub-timestep would find the second derivative of the dependent variable so that the second sub-timestep, the midpoint solution, would only require a three point stencil: the second derivative and initial values at the neighboring spatial points.
The third sub-timestep would find the second derivative using the midpoint solution, and the final sub-timestep would complete the timestep using the results of the second and third sub-timesteps on the three-point stencil and the previous timestep solution at the current spatial point.
This approach presents a storage and data transfer problem on the GPU because values
in the interior of the working array are overwritten two sub-timesteps after they are calculated.
These forgotten but required values are marked with an ``$\times$'' in Figure~\ref{f:secondexes}.

\begin{figure}[!tb]
	\centering
	\begin{subfigure}[t]{.7\textwidth}
		\centering
		\includegraphics[width=\textwidth]{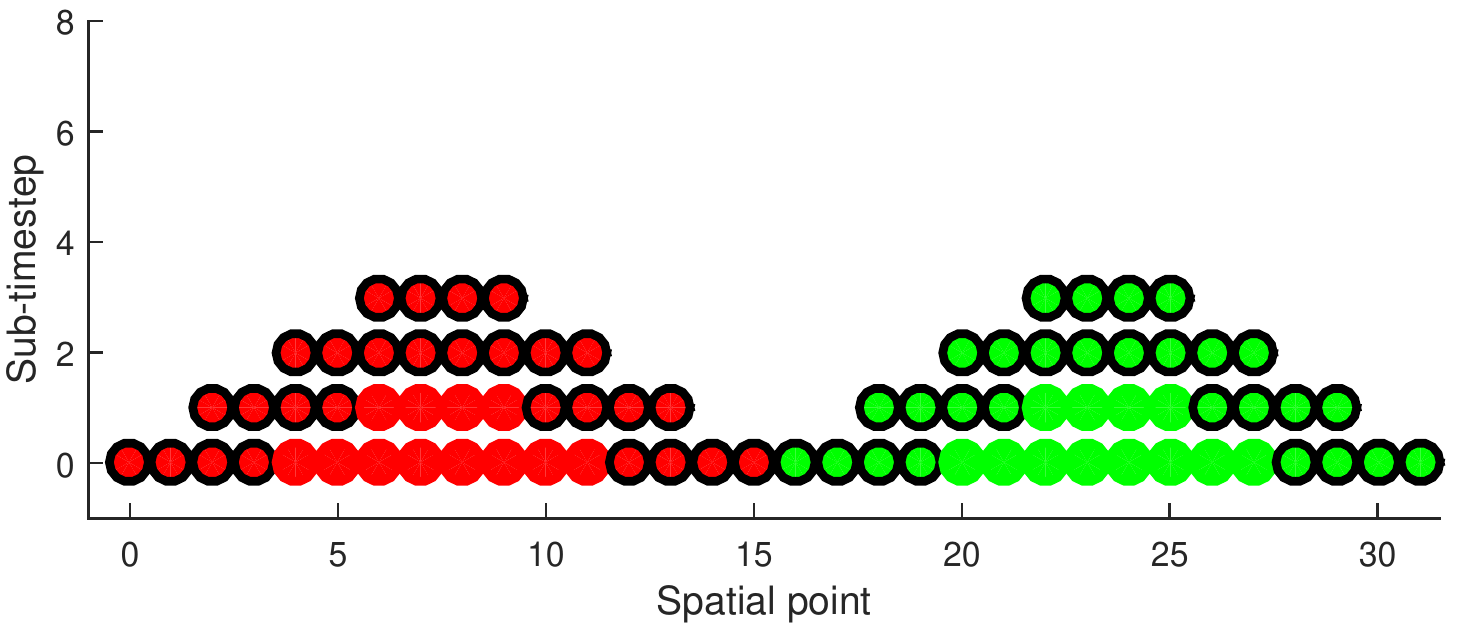}
		\caption{First step in swept rule for second-order domain of dependence.
			The working array is able to be folded and passed as shown in Figure~\ref{f:collection}}
		\label{f:secondsol1}
	\end{subfigure}
	\\
	\begin{subfigure}[b]{.7\textwidth}
		\centering
		\includegraphics[width=\textwidth]{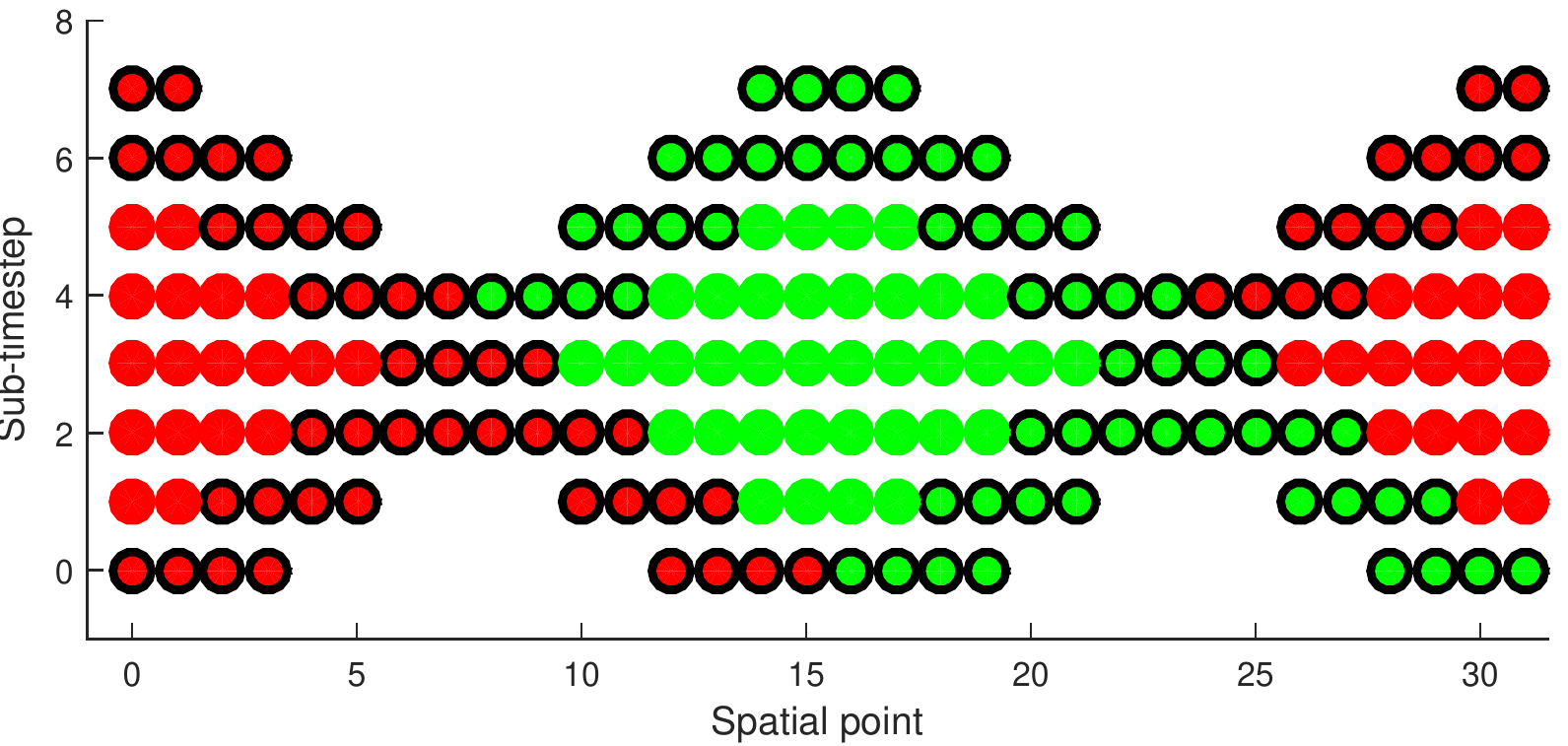}
		\caption{Second step in the swept rule second-order domain of dependence.}
		\label{f:secondsol2}
	\end{subfigure}
	\caption{The first two steps of the swept rule for a numerical scheme with a second-order domain of dependence~\cite{FigsShared}.}
	\label{f:secondordersolution}
\end{figure}

Saving four values per tier would fix the problem, but requires a larger matrix in
shared memory, which would diminish occupancy and require more unnecessary values to be passed between nodes.
Figure~\ref{f:secondordersolution} shows our solution to this problem: a five-point
stencil that requires two sub-timesteps per timestep---a predictor and a final step.
This flattens the triangle or diamond in the time domain and requires four values
per sub-timestep, but the two-row matrix may be used as described in Section~\ref{sec:OrderOne}
and illustrated by Figure~\ref{f:collection} with minor adjustments.
The same number of values are transferred between nodes in each communication, but
inevitably more communication events are required to advance the solution.
Conveniently, the predictor-corrector method ensures that all odd tiers, the second
matrix row, will contain predictor values, and the bottom row will hold final values.

The problem that motivates our adjustment to the swept rule is the result of
both the midpoint method and the five-point stencil discretization.
Either of these circumstances alone would accommodate the method described in the previous section.
Generally, the decomposition must be adjusted in this fashion when there are more
than two sub-timesteps per timestep. It would need to be adjusted further for
problems with more than four sub-timesteps per timestep.

\section{Implementation}\label{sec:schemes}

\subsection{Swept rule variants}

While the order of the swept domain of dependence is a natural consequence of the
equations and numerical scheme, the computational hierarchy available on GPUs
allows this structure to be implemented in different ways.
To more thoroughly investigate the effects of the swept rule, we consider
three versions: \texttt{Shared}, \texttt{Hybrid}, and \texttt{Register}, and compare
them with a basic decomposition algorithm, \texttt{Classic}.
In all of these programs, one GPU thread is assigned to one spatial point.

\begin{figure}[bt]
\begin{minipage}[t]{0.52\textwidth}
\begin{lstlisting}
__global__ void
upTriangle(const REAL *IC, REAL *right, REAL *left) {
//tid = threadIdx.x (bottom row idx),
//tidT = tid+blockDim.x (top row idx);
//tids and tidTs: stencil for each tidT and tid respectively.
//shareT = working array in shared memory
//Read initial data in from IC to shareT.

if (tid > 1 && tid <(blockDim.x-2)) {
	shareT[tidT]=predictorStep(shareT, tids);
	}
__syncthreads();
//4 is stencil length - 1
for (int k = 4; k<(blockDim.x/2); k+=4)
	{

	if (tid < (blockDim.x-k) && tid >= k) {
		shareT[tid]+=finalStep(shareT, tidTs);
		}
	k += 2;
	__syncthreads();
	if(tid < (blockDim.x-k) && tid >= k) {
		shareT[tidT]=predictorStep(shareT,tids);
		}
	__syncthreads();
	}
//Read out the edges to left and right arrays.
}
\end{lstlisting}
	\caption{Main loop of the starting kernel for the swept rule as illustrated by Figure~\ref{f:secondsol1}.}
	\label{f:SweptPseudo}
\end{minipage}
~
\begin{minipage}[t]{0.44\textwidth}
\begin{lstlisting}
__global__ void
classicKS(const REAL *ks_in, REAL *ks_out, bool final)
{
//Global Thread ID
int gid = blockDim.x * blockIdx.x + threadIdx.x;
//number of spatial points - 1
int lastidx = ((blockDim.x*gridDim.x)-1);
//Stencil indices.
int gids[5];

//True for all spatial points from periodic BCs.
#pragma unroll
for (int k = -2; k<3; k++) {
	gids[k+2] = (gid + k) & lastidx;
	}
//Final is false for predictor step, true otherwise.
if (final) {
	ks_out[gid] += finalStep(ks_in, gids);
	}
else {
	ks_out[gid] = predictorStep(ks_in, gids);
	}
}
\end{lstlisting}
	\caption{\texttt{Classic} kernel for Kuramoto-Sivashinsky solver. Final and predictor step functions can be written in C with \texttt{\_\_device\_\_} keyword.}
	\label{f:ClassicCode}
\end{minipage}
\end{figure}

The \texttt{Classic} algorithm is a naive GPU implementation of the numerical solution
and the baseline against which the efficacy of the swept rule is measured.
It advances one sub-timestep per kernel call and uses global memory to store the working array.
Figure~\ref{f:ClassicCode} shows the structure of the procedure, and it is
similar for all problems and discretizations.

We consider the \texttt{Shared} strategy for implementing the swept rule
the most natural way to map the analogy of CPU nodes to GPU architecture.
It is applied to every test case and is considered the ``default'' swept
rule GPU version for comparing the performance of the GPU programs with their MPI-based
CPU counterparts~\cite{MaithamRepo}. The \texttt{Shared} version treats a block as a node
and uses shared memory for the working array.
Each block has exclusive access to a shared memory space and may contain up to
\num{1024} threads, so it has access to fast memory and the capacity for various node sizes.
There are some drawbacks to this version: a high number of idle threads for sub-timesteps
where the domain of dependence contains relatively few spatial points, poor utilization
of CPU resources, and the fact that shared memory is not the fastest memory type~\cite{harris:2014}.

The \texttt{Hybrid} strategy uses the same GPU procedure as \texttt{Shared}, uses the CPU to compute the node that is split across the boundary, as seen in Figure~\ref{f:firstorder2},
Transfers between the host and device are costly operations, but the devices can
execute instructions concurrently---so if the CPU can complete the boundary node
before the GPU finishes the other nodes, no penalty arises.
In this study, we apply this strategy to problems with non-periodic boundary conditions
as a way to mitigate the underutilization of the CPU and the thread divergence that
results from applying boundary conditions in a GPU kernel.

The \texttt{Register} approach is applied to problems with periodic boundary conditions
because of the difficulty involved in applying boundary conditions using warp shuffle functions.
This implementation limits the number of points in a node to the size of a warp, \num{32}
threads (which has been constant over several iterations of NVIDIA GPUs).
In this version the values are initially read from global memory to shared memory to the registers.
Passing the values to the intermediate shared memory is necessary because the shuffle
operations that trade registers between threads only operate on active threads in the
warp being called; if some threads are masked, they will be unable to supply the necessary stencil values.
Thus, data must be moved between memory levels at each tier rather than once at the start and end of the kernel.
This seriously limits the \texttt{Register} approach, but it still warrants exploration
since registers are the fastest memory type.

\subsection{Test cases} \label{sec:tests}

We present three test cases to demonstrate the performance and functionality of the
GPU-based swept rule in one spatial dimension: the heat equation, Kuramoto--Sivashinsky (KS)
equation, and Euler equations for compressible flow.
Appendices~\ref{app:heat_eq}--\ref{app:euler} contain the full derivations of the
procedures for the numerical solutions.
First, we chose the heat equation for its simplicity and familiarity.
Here it is discretized with a first-order scheme using forward differencing in
time and central in space. Next, we selected the KS equation to demonstrate the
swept rule for higher-order, nonlinear PDEs.
We discretized the KS equation with second-order, central differencing in space,
which requires a five-point stencil, and a second-order Runge--Kutta method in time.
Lastly we chose to solve the Euler equations, a system of quasilinear, hyperbolic
equations for describing compressible, inviscid flow.
The conservative form of these equations are applied to the Sod shock tube problem to
demonstrate the application of the swept rule to a canonical CFD problem involving
discontinuities and several dependent variables. These equations are discretized
with a second-order, finite-volume scheme in space and a second-order method in time.
The heat equation requires a first-order domain of dependence, while the
KS and Euler equations require second-order domains of dependence.
These problems also provide examples of various types of boundary conditions.
The heat and Euler equations are solved with reflective and Dirichlet boundary conditions,
respectively; therefore, the boundary conditions must be imposed with control flow.
The KS equation uses periodic boundary conditions, which is a convenient formulation
for the swept rule that splits a node across the boundary.

\section{Results and discussion} \label{sec:results}

All tests presented here were performed on a single workstation with a Tesla K40c GPU
and an Intel Xeon 2630-E5 CPU with eight cores and 16 potential threads.
The GPU-based swept rule algorithms and test cases were implemented \texttt{1DSweptCUDA v2}~\cite{MyRepo}.
For the results we present here, each program was executed in double precision with
$\{2^x \mid x \in \mathbb{N} \mid 4<x<11\}$ threads per block and
$\{2^x \mid x \in \mathbb{N} \mid 10<x<21\}$ spatial points.
Each run advanced \num{50000} timesteps and recorded the average time per timestep;
the initial GPU memory allocations and data transfer between the host and device are included in the overall time measurement.
Then, we collected the best time per timestep for each number of spatial points.
We repeated this procedure five times and took the average to obtain the results presented here.
There were no significant differences in the results between the tests for the same configuration.

\begin{figure}[!tb]
	\centering
	\begin{subfigure}[t]{.48\textwidth}
		\centering
		\includegraphics[width=\textwidth]{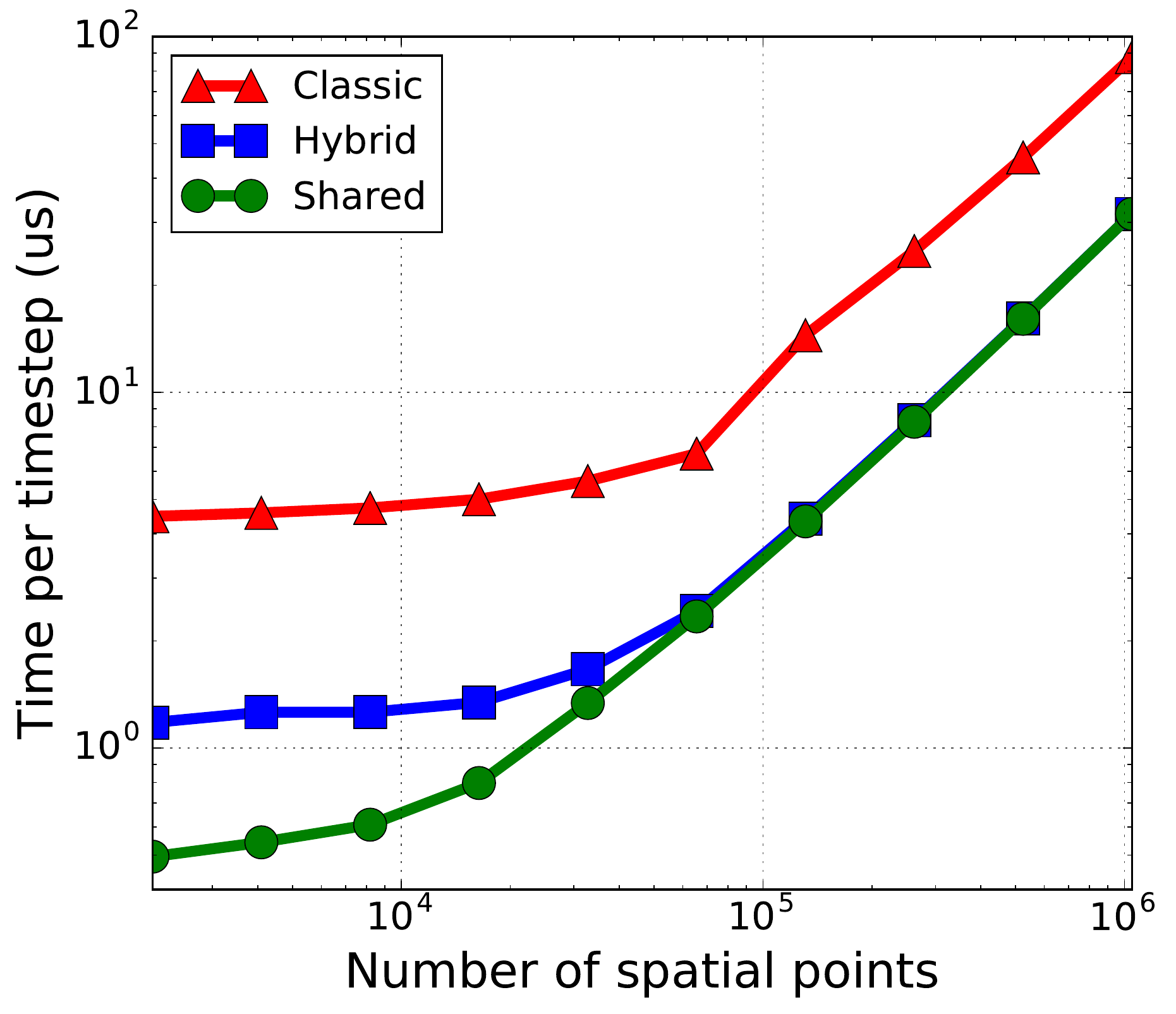}
		\caption{Time cost of GPU classic and swept domain decomposition algorithms.}
		\label{f:HeatRaw}
	\end{subfigure}
	\hfill
	\begin{subfigure}[t]{.48\textwidth}
		\centering
		\includegraphics[width=\textwidth]{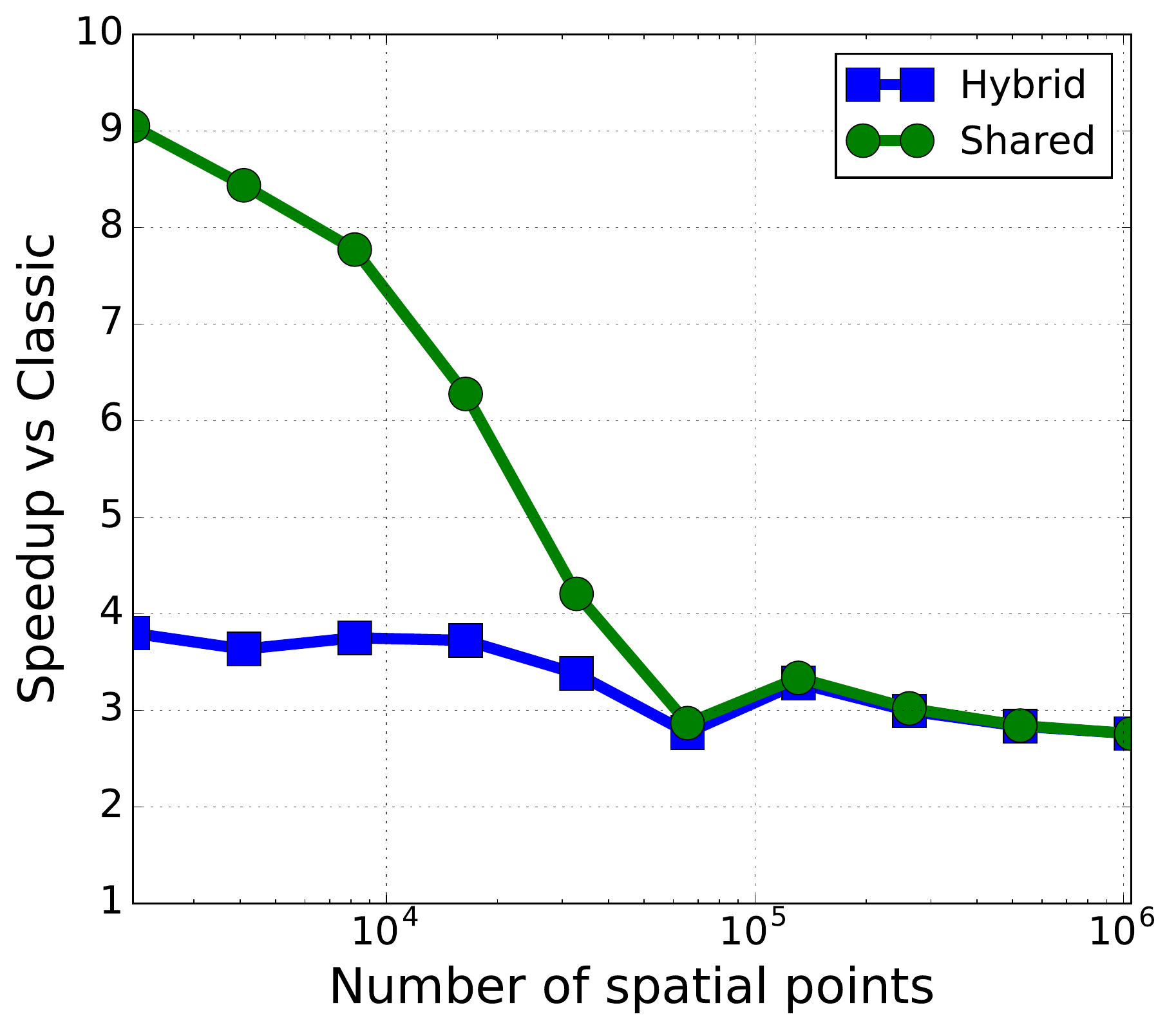}
		\caption{Speedup of swept rule programs with respect to the \texttt{Classic} version.}
		\label{f:HeatSpeed}
	\end{subfigure}
	\caption{Performance comparison of the GPU heat equation programs~\cite{FigsShared}.}
	\label{f:HeatResult}
\end{figure}

\begin{figure}[!tb]
	\centering
	\begin{subfigure}[t]{.48\textwidth}
		\centering
		\includegraphics[width=\textwidth]{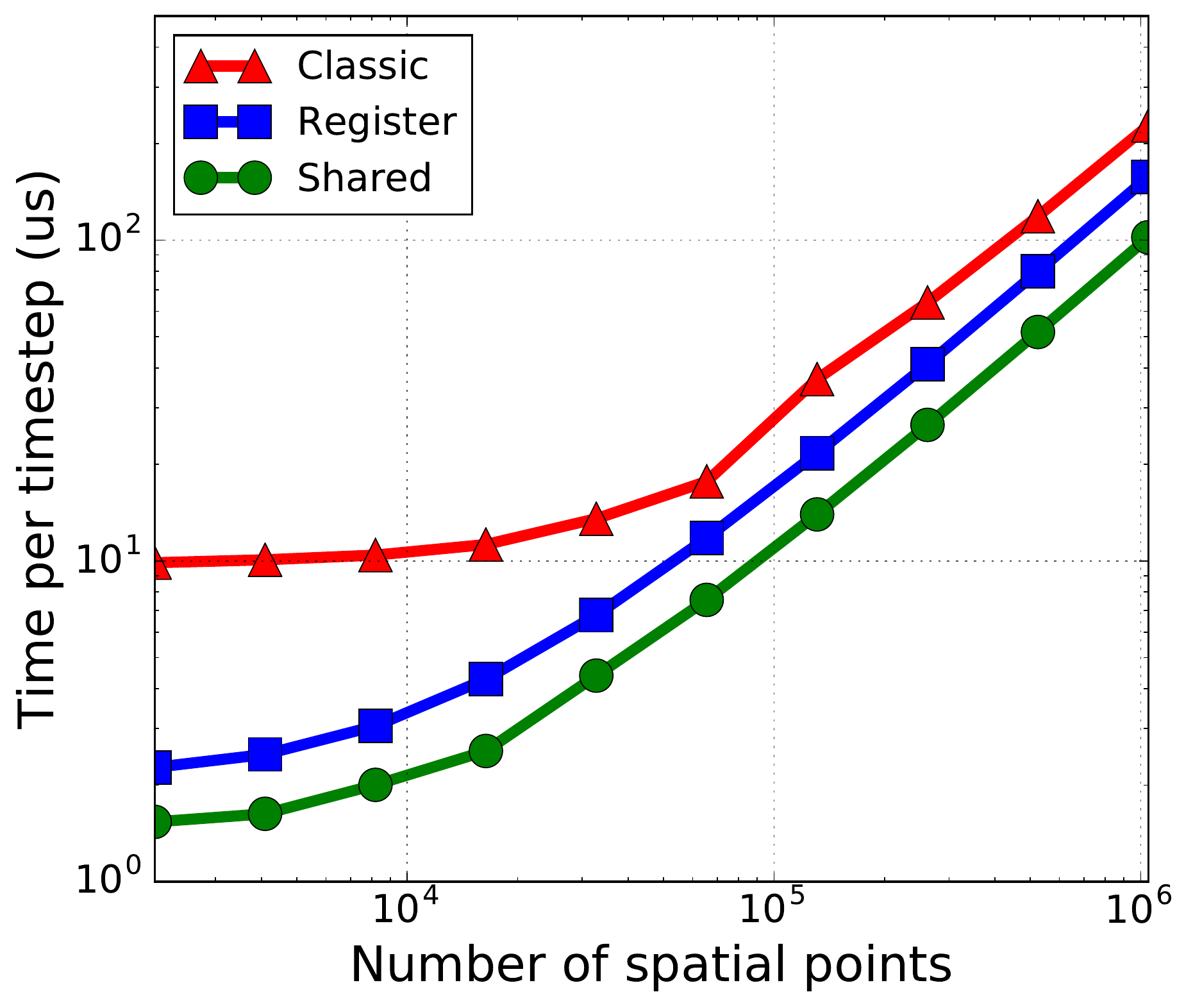}
		\caption{Time cost of GPU classic and swept domain decomposition algorithms.}
		\label{f:KSRaw}
	\end{subfigure}
	\hfill
	\begin{subfigure}[t]{.48\textwidth}
		\centering
		\includegraphics[width=\textwidth]{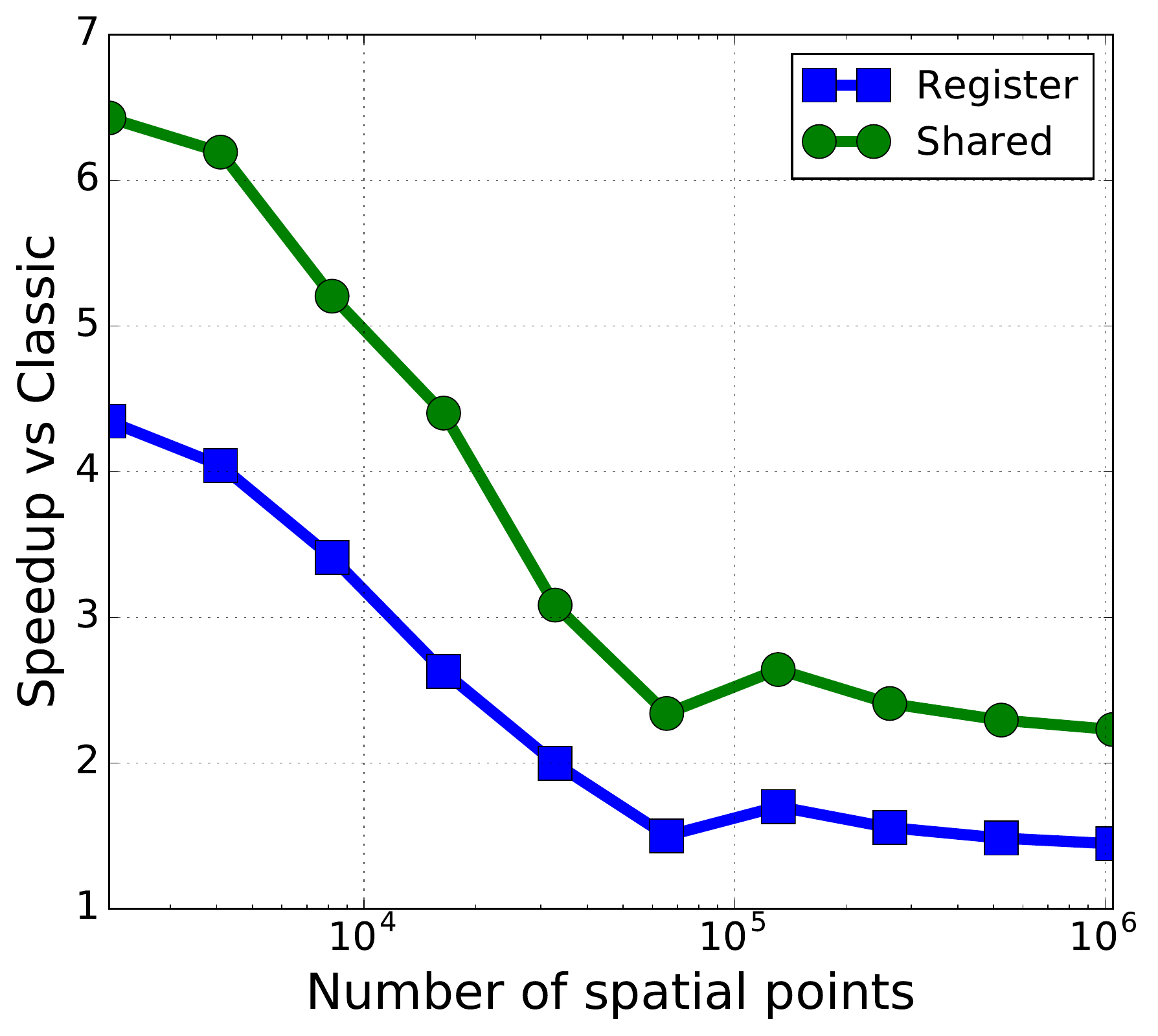}
		\caption{Speedup of swept rule programs with respect to the \texttt{Classic} version.}
		\label{f:KSSpeed}
	\end{subfigure}
	\caption{Performance comparison of the GPU Kuramoto--Sivashinsky equation programs~\cite{FigsShared}.}
	\label{f:KSResult}
\end{figure}

Figures~\ref{f:HeatRaw} and~\ref{f:KSRaw} show the execution time per timestep for all algorithms
applied to the heat and KS equations. When applied to these problems, the swept rule
algorithm outperforms the \texttt{Classic} procedure, and the GPU-only shared memory
version, \texttt{Shared}, is faster than the alternate swept rule versions.
Figures~\ref{f:HeatSpeed} and~\ref{f:KSSpeed} show the speedup of the swept rule programs, or
the ratio of time costs with the \texttt{Classic} version.
Both cases exhibit similar performance patterns: \texttt{Shared} generally provides a
larger speedup for small spatial domains ($< \num{e4}$ spatial points), but only a
\num{2}$\times$ speedup for large ones ($> 10^5$ spatial points).
Figures~\ref{f:HeatRaw} and \ref{f:KSRaw} show the performance trends of the algorithms
with respect to the spatial domain size.
Their time costs are insensitive to the spatial domain at smaller domain sizes and grow
linearly with increasing domain size after about \num{3e5} spatial points.
The similarity in the performance trends of the heat and KS programs is intuitive;
although the KS equation is nonlinear, fourth-order, and discretized with a
higher-order scheme, both are continuous, scalar equations and use finite difference schemes.


The speedup of the swept rule declines as the number of spatial points in the domain increases,
caused primarily by occupancy and the fixed cost of kernel launch.
The occupancy of a kernel is the number of threads that can reside concurrently on a streaming
multiprocessor at launch. This quantity is determined by the block size, since blocks of
threads are indivisible, and the resources requested by the kernel.
The swept rule requests more resources, specifically shared memory and registers, than
\texttt{Classic} to store the working array and carry out its procedure, which
involves more steps. If more resources are requested than are available to the streaming
multiprocessors on the device, the GPU launches waves of blocks.
If the occupancy is limited by resource allocation, the kernel must begin launching waves
of blocks on domains with fewer spatial points.
Each wave must wait until the previous one completes before beginning; theoretically this
implies two waves will cost about twice as much as one.
As the number of spatial points in the domain grows, more blocks are launched, and the
difference in the number of waves per kernel call increases more quickly for the swept rule program.
This conclusion arises from the observation that the time cost of the swept rule versions
begins growing linearly at a smaller spatial domain size, about \num{2e4} points, than
\texttt{Classic}, about \num{6e4} points, as shown in Figures~\ref{f:HeatRaw} and~\ref{f:KSRaw}.

By timing the launch of many empty kernels and taking their average, we measured the fixed
cost of kernel launch on the testing workstation to be about \SI{4}{\micro\second}.
We conclude that this cost dominates the performance of \texttt{Classic} at small
problem sizes because it is quite close to the cost of each timestep for spatial
domains with less than \num{3e4} spatial points for both the heat and KS equations.
This fixed cost accounts for less time cost at larger spatial domain sizes where all
kernels must launch several waves of blocks, so the portion of the swept rule speedup
from avoiding kernel launches becomes negligible, and the overall speedup of the swept
rule is reduced.

Each of the swept rule variants experiences these trends, but in all cases the \texttt{Shared} version performs better.
Figure~\ref{f:HeatResult} shows the heat equation test, where the \texttt{Hybrid} version performs as well as
\texttt{Shared} for large spatial domain sizes and 2--3 times worse for small ones.
The \texttt{Hybrid} version uses the same GPU kernels as \texttt{Shared}, but uses the CPU to
calculate the first node when it is split across the boundary.
At smaller domain sizes, the data transfer between the host and device dominates the
cost of the \texttt{Shared} scheme. At larger domain sizes, this cost drops in
comparison with the overall costs, but at the same time the thread divergence
caused by handling boundary conditions also drops in importance.
Thus, the \texttt{Shared} and \texttt{Hybrid} methods perform similarly at larger
domain sizes (i.e., above \num{6e4} spatial points).

Figure~\ref{f:KSResult} shows the KS equation test, where, similar to the heat equation,
the \texttt{Shared} version performs better in all cases.
While registers are the fastest memory type, because of memory access rules the
\texttt{Register} version must use a warp as a node, which limits the domain of
dependence to 32 spatial points.
A \texttt{Register} program with $n=32$ must launch four times as many kernels to
advance the same number of timesteps as \texttt{Shared} with $n=128$ (most often
the best block size).
Despite these differences, the \texttt{Register} version exhibits a similar performance
trend to \texttt{Shared}, which shows about \num{1.5}$\times$ the speedup of
\texttt{Register} at all spatial domain sizes.
The limit on node size results in a constant four timesteps per cycle for the KS
equation, which only reduces the number of communication events by $1/8$ compared
with \texttt{Classic}.

\begin{figure}[!tb]
	\centering
	\begin{subfigure}[t]{.48\textwidth}
		\centering
		\includegraphics[width=\textwidth]{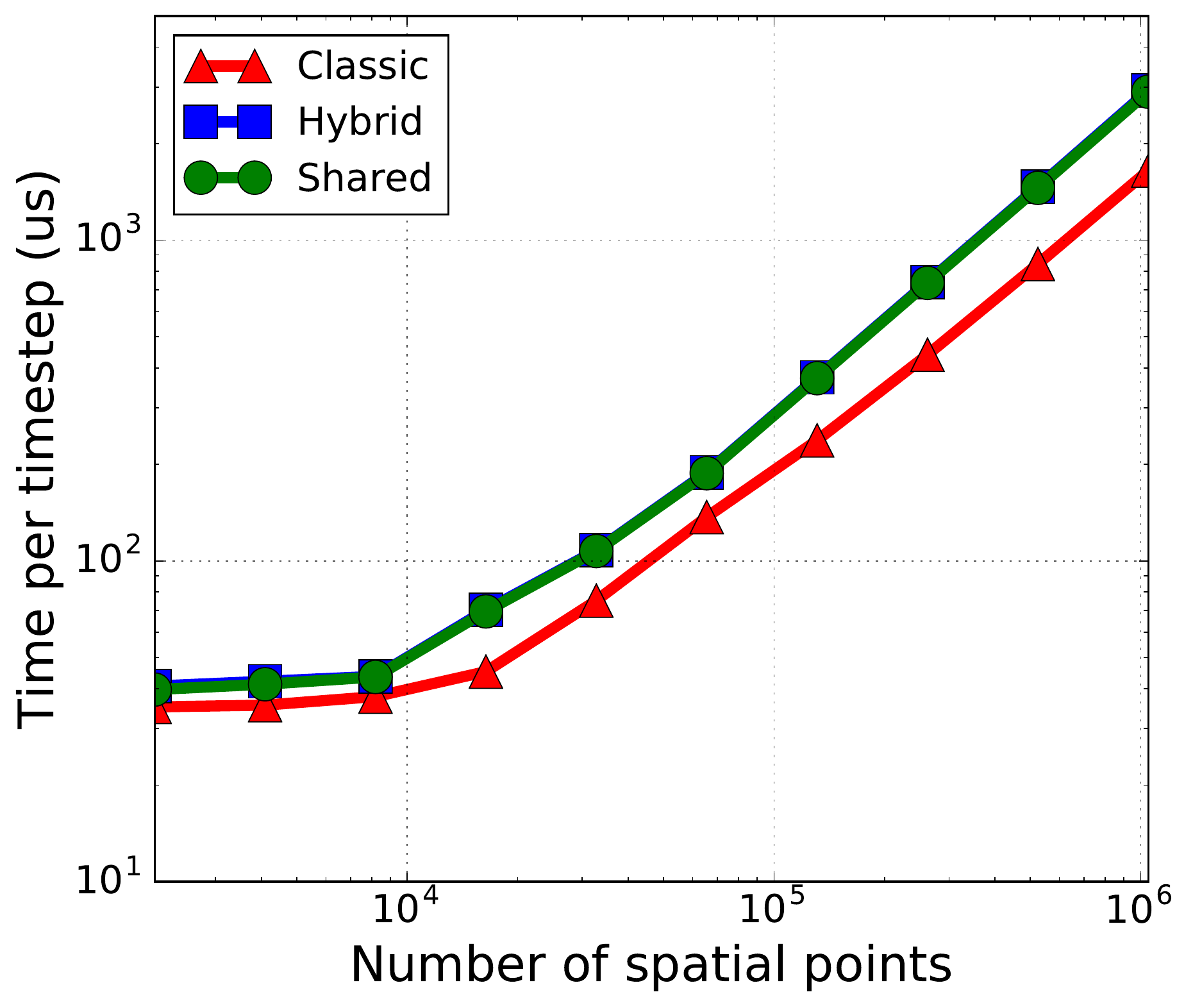}
		\caption{Time cost of GPU classic and swept domain decomposition algorithms.
        \texttt{Hybrid} and \texttt{Shared} lines overlap.
        }
		\label{f:EulerRaw}
	\end{subfigure}
	\hfill
	\begin{subfigure}[t]{.48\textwidth}
		\centering
		\includegraphics[width=\textwidth]{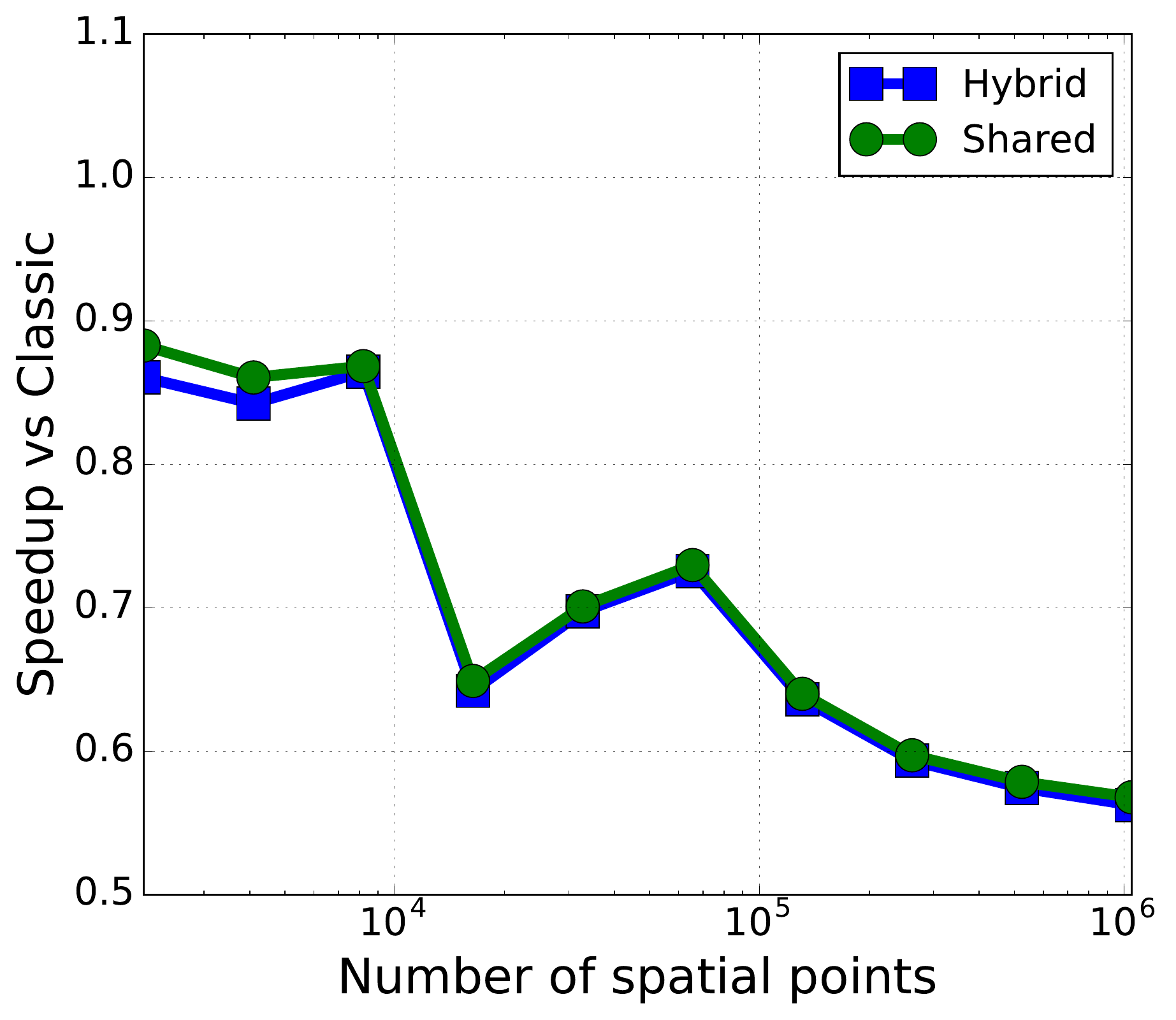}
		\caption{Speedup of swept rule programs with respect to the \texttt{Classic} version.}
		\label{f:EulerSpeed}
	\end{subfigure}
	\caption{Performance comparison of the GPU Euler equation programs~\cite{FigsShared}.}
	\label{f:EulerResult}
\end{figure}

In contrast to the previous test cases, Figure~\ref{f:EulerResult} shows that the classic
decomposition outperforms the swept rule at all problem sizes when applied to the Euler equations.
This result differs from the findings of Alhubail and Wang~\cite{alhubail:16jcp} for
the swept rule implemented only with MPI on CPUs, which produces roughly equivalent speedups
for both the KS and Euler equations.
The GPU implementation of the swept rule for the Euler problem involves much
greater arithmetic intensity than the other problems, causing greater
low-level memory usage. This limits the performance of the swept scheme on a GPU
more than a CPU, reducing the benefits of the scheme over the classic approach
within a single GPU compared with the other problems. Those cases involve higher
ratios of floating-point operations to memory accesses (both read and write).
This will particularly degrade performance as the intra-domain timesteps proceed
and nodes become inactive.

With regard to the potential performance of the swept rule, this result is not
encouraging for more complex problems in more dimensions. However, on a larger scale
in a cluster, where global memory is used for the working array and the communication
to be avoided is through a physical network between processors, we expect
that the swept rule will provide a benefit for these types of problems.

\begin{figure}[!tb]
	\centering
	\begin{subfigure}[t]{.48\textwidth}
		\centering
		\includegraphics[width=\textwidth]{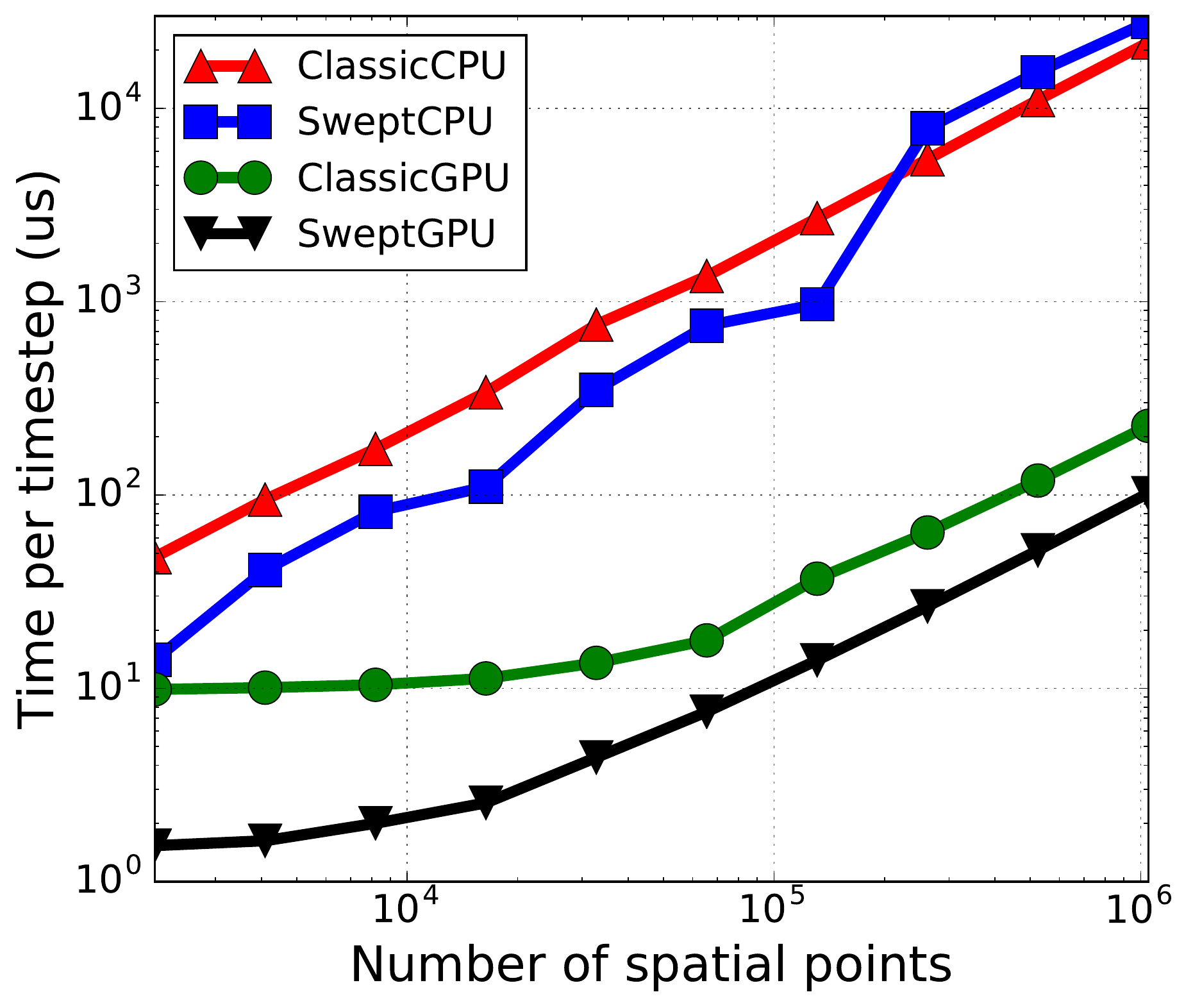}
		\caption{Time cost of CPU and GPU programs for KS equation.}
		\label{f:mpiraw}
	\end{subfigure}
	\hfill
	\begin{subfigure}[t]{.48\textwidth}
		\centering
		\includegraphics[width=\textwidth]{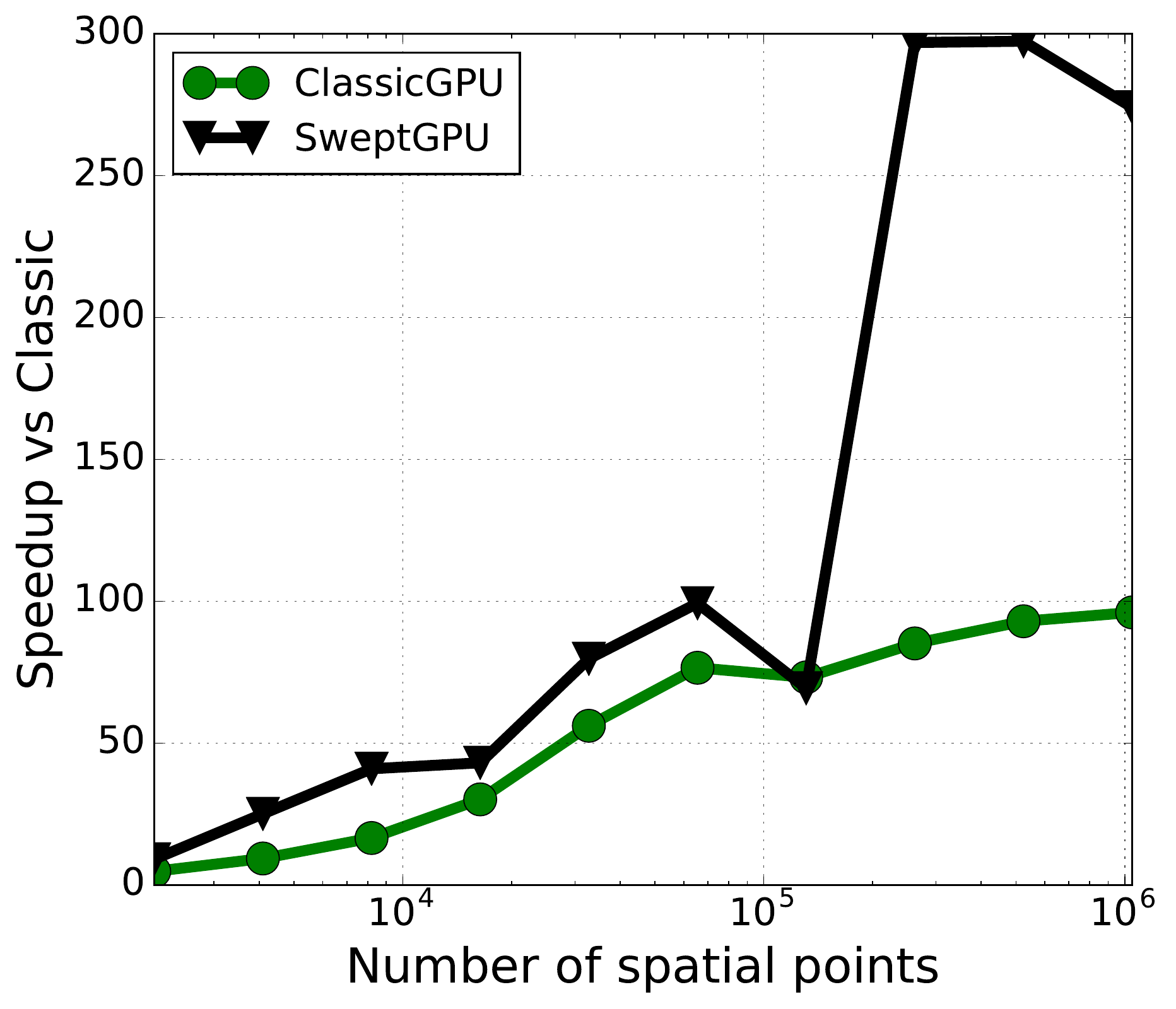}
		\caption{Performance improvement of GPU program compared to the same algorithm on the CPU.}
		\label{f:mpispeed}
	\end{subfigure}
	\caption{Performance comparison of CPU (MPI) and GPU (CUDA) programs for the KS equation~\cite{FigsShared}.}
	\label{f:mpi}
\end{figure}

Figure~\ref{f:mpi} compares the CPU-based KS equation programs with the GPU algorithms,
\texttt{Classic} and \texttt{Shared}.
This CPU version is parallelized with MPI, similar to that originally presented by
Alhubail and Wang~\cite{MaithamRepo}.
Figure~\ref{f:mpi} refers to the \texttt{Shared} program as \texttt{SweptGPU} because it serves
as the ``default'' swept rule version on the GPU for comparison with the CPU-based swept rule.
The original CPU program was designed for a distributed-memory cluster, but the
Alhubail and Wang's original performance study used
only two processes on two CPUs~\cite{alhubail:16jcp}.
The parallel CPU program was tested similarly to the GPU programs: it was
run on the same workstation with the same spatial domain sizes and \numrange{2}{16} threads.
Here we compare the best results for each number of spatial points, which usually
occurred with 16 threads. This did not degrade the performance compared with the original study;
in fact, both CPU versions of the program performed significantly better, and
for most spatial domain sizes the CPU-based swept rule improved by about \SI{6}{$\times$}.
Since the test used up to eight times the number of threads, this result
supports the validity of repurposing this code and comparing the result with the GPU versions.

Figure~\ref{f:mpi} illustrates the performance benefits of the GPU architecture more than the swept rule itself.
On small spatial domains (e.g., less than \num{1e4} points) the GPU can assign one
thread to each spatial point and process all in a single wave.
There it improves performance over the CPU version by less than \SI{50}{$\times$}, because the
work does not fully utilize the GPU. On larger spatial domains (e.g., more than \num{1e4}
points), where the work completely utilizes the resources of the GPU, the performance
increases continue growing with problem size.
The increase of the GPU programs' improvement with respect to the number of points in the spatial domain
inverts the trend of the swept rule's improvement with respect to the \texttt{Classic} kernel.
Both GPU algorithm types show a speedup of about \SI{5}{$\times$} for the smallest spatial domain.
At the other end, the speedup grows to about \SI{300}{$\times$} for the swept and
\SI{100}{$\times$} for the classic domain decomposition.

\section{Conclusions} \label{sec:conclude}

In this study we compared the time per timestep of three swept rule time-space decomposition
implementations to a simple domain decomposition scheme, \texttt{Classic}, for GPU accelerators.
These programs were evaluated on a range of spatial domain sizes.
The \texttt{Classic} and \texttt{Shared} programs were also compared with their
CPU-based counterparts parallelized with MPI.

Generally, the swept rule performs better relative to \texttt{Classic} when the
kernel launch cost is a substantial element of the program cost and the spatial
grid has fewer points than concurrently launchable threads on the GPU.
Once this is exceeded, the launch cost penalty becomes negligible and greater resource
usage penalizes the swept rule in the form of reduced occupancy and reduced availability
of the L1 cache, which is reserved for shared memory.
But, like the initial performance penalty for the \texttt{Classic} program, the
resource usage penalty does not scale and all programs see their time cost rise
nearly linearly with respect to spatial domain size after about~\num{e5} spatial points.

Both alternate swept-rule procedures, \texttt{Hybrid} and \texttt{Register}, performed
slightly worse than the \texttt{Shared} version. We attribute the failure of the \texttt{Hybrid} routine
to improve performance to improvements made in the handling of the boundary conditions in \texttt{Shared}
rather than the cost of host-to-device communication, which is handled with asynchronous streams.
Though the \texttt{Hybrid} version does not improve performance, it does not substantially
degrade it either, especially with large spatial domain sizes.
Ultimately, \texttt{Hybrid} computation seeks to solve a problem more efficient to solve
within the GPU paradigm. The \texttt{Register} computation shows more promise but
falls short because of the limited node size.

This study also shows that implementing a \texttt{Classic} decomposition on the
GPU for an explicit numerical scheme is simple and may result in noticeable performance
improvements. This may be enough for many applications, but in performance-critical cases
the swept rule may further reduce execution time.
The performance of the \texttt{Classic} decomposition programs may be improved by altering the synchronization method.
Implicit synchronization incurs the cost of kernel launch at each sub-timestep, and
this kernel executes so quickly at small problem sizes that synchronization cost
dominates the performance of the program.
Using off-the-shelf, GPU-based synchronization~\cite{GlobalLock} could also provide
performance benefits for the swept rule, and in particular \texttt{Register}, which is
also limited by kernel launch cost for small spatial domains.

The specific results presented here depend on the hardware used.
A commercial GeForce GPU with Kepler architecture would perform worse than the Tesla K40c,
which is designed for computation as opposed to actual graphics.
Despite the significant performance increase shown here, the Tesla K40c is three-year-old technology.
The current state-of-the-art GPGPU with Pascal architecture, the Tesla P100, offers
twice the K40c base clock speed, nearly four times the number of streaming
multiprocessors, and an extra \SI{16}{\kilo\byte} of shared memory independent of the L1 cache.
Additional dedicated shared memory could dramatically impact the swept rule's performance for the Euler equations.
We predict that this device could at least halve the execution times shown here and
maintain insensitivity to problem size up to over \num{e5} spatial points,
potentially resulting in speedups on the order of \num{e3} over CPU parallel versions.
Future GPU accelerators could further improve performance, as well as devices with
similar SIMT architectures like Intel MIC\slash Xeon Phi.

Our future work will focus on further developing the swept rule for use on distributed
memory systems with heterogeneous node architectures and
implementing the swept rule in higher dimensions on the GPU.
We expect the performance benefits to increase from the swept rule when applied
to a distributed memory system comprised of nodes with CPUs and GPUs, where network
communications incur more latency than intra-GPU memory access.
However, we anticipate the performance benefits of the swept rule will diminish for higher dimensions;
in two dimensions the swept rule requires three stages of inter-node communication to
advance one cycle, analogous to a single diamond in one dimension.
This, along with the associated increase in required memory allocation, arithmetic
intensity, and kernel launch events could limit the performance of the swept
rule as observed by Alhubail et al.~\cite{Alhubail:2016arxiv}. In spite of these challenges,
recent developments such as grid-wide synchronization and increased shared memory
capacity could aid the performance of the swept rule and permit the exploration
of additional design options.

\appendix

\renewcommand*{\thesection}{\appendixname~\Alph{section}}

\label{app:availability}
\section{Availability of materials}
The results for this article were obtained using \texttt{1DSweptCUDA} v2~\cite{MyRepo}.
All figures, and the data and plotting scripts necessary to reproduce them, are
available openly under the CC-BY license~\cite{FigsShared}.

\section{Heat equation} \label{app:heat_eq}
The unsteady heat conduction equation without volumetric heat flux is
\begin{equation}
\frac{\partial T}{\partial t} = \alpha \nabla^2 T \;.
\label{eq:Heatfull}
\end{equation}
where $T$ is temperature, $t$ is time, and $\alpha$ is thermal diffusivity.
In one dimension, this is reduced to
\begin{equation}
\frac{\partial T}{\partial t} = \alpha \frac{\partial^2 T}{\partial x^2} \;.
\label{eq:Heat1d}
\end{equation}

Discretizing Eq.~\eqref{eq:Heat1d} with forward differencing in time and central in space yields
\begin{equation}
\frac{T_i^{m+1}-T_i^m}{\Delta t} = \alpha \frac{T_{i+1}^m + T_{i-1}^m + 2T_i^m}{\Delta x^2} \;,
\end{equation}
where $i$ is the spatial node index and $m$ is the time index corresponding to time $t^m$.
This is a first-order, explicit, finite-difference approximation.
To step forward in time, define the Fourier number, $\text{Fo} = \frac{\alpha \Delta t}{\Delta x^2} $, where $\Delta t$ is the timestep size and $\Delta x$ is the spatial grid size, and solve for temperature at the next timestep:
\begin{equation}
T_i^{m+1} = \text{Fo} (T_{i+1}^m + T_{i-1}^m) + (1 - 2 \text{Fo}) T_i^m \;.
\label{eq:Heatdisc}
\end{equation}
In this study the approximation is evaluated with insulated boundary conditions at both ends and $n$ spatial points:
\begin{equation}
T_{-1} = T_1 \quad \text{and} \quad T_{n+1} = T_{n-1} \;.
\label{eq:heatcond}
\end{equation}

\section{Kuramoto--Sivashinsky equation}\label{app:KS}
The Kuramoto--Sivashinsky equation is a nonlinear, fourth-order, one-dimensional unsteady PDE:
\begin{equation}
u_t = -(uu_x + u_{xx} + u_{xxxx})
= -\left( \frac{1}{2} u_x^2 + u_{xx} + u_{xxxx} \right) \;,
\label{eq:KSfull}
\end{equation}
where $u$ is the dependent chaotic variable (e.g., fluid velocity).
It is discretized similarly to the heat equation, as shown in Eq.~\eqref{eq:KSDisc}, with central differencing in space.
In this case, decomposing the fourth spatial derivative requires a five-point stencil:
\begin{multline}
\frac{u_i^{m+1}-u_i^m}{\Delta t}
= -\left( \frac{(u_{i+1}^m)^2 - (u_{i-1}^m)^2}{4\Delta x} +
\frac{u_{i+1}^m + u_{i-1}^m + 2u_i^m}{\Delta x^2} + \right. \\
\left. \frac{u_{i+2}^m - 4u_{i+1}^m + 6u_i^m - 4u_{i-1}^m + u_{i-2}^m}{\Delta x^4} \right) \;.
\label{eq:KSDisc}
\end{multline}

The chaotic nature of the problem necessitates a higher-order scheme overall; therefore, an explicit, second-order Runge--Kutta scheme, also known as the midpoint method, is applied to the time domain.

Let the right-hand side of Eq~\eqref{eq:KSDisc} be $f(u(x,t))$, then the predictor solution is found at $u_i^{m+1/2}$:
\begin{equation}
u_i^{m+1/2} = u_i^m + \frac{\Delta t}{2} f(u_i^m) \;.
\label{eq:RKm}
\end{equation}

Then $f(u_i^{m+1/2})$ may be evaluated and added to $u_i^{m}$ to obtain $u_i^{m+1}$:
\begin{equation}
u_i^{m+1} = u_i^m + \Delta t f(u_i^{m+1/2}) \;.
\label{eq:RKf}
\end{equation}

The problem demonstrated here uses periodic initial and boundary conditions.
That is, the stencil at point $0$ includes points $n$ and $n-1$ and the initial condition
is periodic and continuous at the spatial boundaries.

\section{Euler equations (Sod shock tube)} \label{app:euler}
The Sod shock tube problem is a one-dimensional unsteady compressible flow problem
based on the nonlinear, quasi-hyperbolic Euler equations:
\begin{equation}
\frac{\partial Q}{\partial t} + \frac{\partial F}{\partial x} =
\frac{\partial Q}{\partial t} + J\frac{\partial Q}{\partial x} = 0
\label{eq:EulerGeneral}
\end{equation}
where $J$ is the Jacobian matrix,
\begin{align}
Q &= \begin{Bmatrix}
\rho \\ \rho u \\ \rho e
\end{Bmatrix} \;,
\label{eq:EulerQ} \\
F &= \begin{Bmatrix}
\rho u \\ \rho u^2 + P \\ u(\rho e + P)
\end{Bmatrix} \;,
\label{eq:EulerF}
\end{align}
$\rho$ is density, $e$ is internal energy, $u$ is velocity, $P$ is pressure given by
\begin{equation}
P = (\gamma - 1)(e-\frac{\rho u^2}{2}) \;,
\label{eq:EulerP}
\end{equation}
and $\gamma=1.4$ is the heat capacity ratio of air.

The initial boundary conditions, given in Eq.~\eqref{eq:SodCond}, are constant values for the state variables on either side of a diaphragm separating two parcels of the same fluid. The spatial boundary conditions are these values at their respective ends of the tube:
\begin{equation}
\begin{Bmatrix}
\rho_L \\ u_L \\ P_L
\end{Bmatrix}
=
\begin{Bmatrix}
1.0 \\ 0.0 \\ 1.0
\end{Bmatrix}
\text{ and }
\begin{Bmatrix}
\rho_R \\ u_R \\ P_R
\end{Bmatrix}
=
\begin{Bmatrix}
0.125 \\ 0.0 \\ 0.1
\end{Bmatrix} \;.
\label{eq:SodCond}
\end{equation}

The equation is discretized using a second-order, finite-volume scheme with cell-centered values. The first step in the solution is evaluating the pressure ratio at the current timestep over a five-point stencil
\begin{equation}
P_{r,i} = \frac{P_{i+1}-P_i}{P_i-P_{i-1}}
\;\; \text{at} \;\; i-1,\;i,\;i+1
\label{eq:Pratio}
\end{equation}

This value is used with a minmod limiter to compute reconstructed values on both sides, $L$ and $R$, of the current cell boundaries at $i\pm1/2$:
\begin{equation}
Q_{n}^{L} =
\begin{cases}
Q_{o}^{L} + \frac{\min(P_{r}^{L}, 1)}{2} * (Q_{o}^{R}-Q_{o}^{L}), &
\text{if } 0 < P_{r}^{L} < \infty \\
Q_{o}^{L}, & \text{otherwise.}
\end{cases}
\label{eq:limitleft}
\end{equation}
\begin{equation}
Q_{n}^{R} =
\begin{cases}
Q_{o}^{R} + \frac{\min((P_{r}^{R})^{-1}, 1)}{2} * (Q_{o}^{L}-Q_{o}^{R}), &
\text{if } 0 < \frac{1}{P_{r}^{R}} < \infty \\
Q_{o}^{R}, & \text{otherwise.}
\end{cases}
\label{eq:limitright}
\end{equation}
where subscript $n$ refers to the reconstructed, or new, values on the edge of the interface and $o$ refers to the original values. For example, at $i-1/2$ the original values on the left side of the interface are at $i-1$.
These reconstructed boundaries do not represent solutions for any grid cell; they are temporary values that interpolate the solutions.

Once we have the reconstructed values on either side of the interface, we can
calculate the flux at that cell boundary with
\begin{equation}
\text{Flux} = \frac{1}{2}*(F(Q^R)+F(Q^L) + r_{sp} * (Q^{L} - Q^{R}))
\end{equation}
where $F(Q)$ is given by Eq.~\eqref{eq:EulerF} and $r_{sp}$ is the spectral radius, the largest eigenvalue of the Jacobian matrix $J$.

The spectral radius can be found with the Roe average $Q$ at the interface
\begin{equation}
Q_{sp} = \begin{Bmatrix}
\rho_{sp} \\
u_{sp} \\
e_{sp}
\end{Bmatrix}
=
\begin{Bmatrix}
\sqrt{\rho_{L}*\rho_{R}} \\
\frac{\sqrt{\rho_{L}}*u_{L}+\sqrt{\rho_{R}}*u_{R}}{\sqrt{\rho_{L}}+\sqrt{\rho_{R}}} \\
\frac{\sqrt{\rho_{L}}*e_{L}+\sqrt{\rho_{R}}*e_{R}}{\sqrt{\rho_{L}}+\sqrt{\rho_{R}}}
\end{Bmatrix}
\end{equation}
and $P_{sp}$ with $Q_{sp}$ using Eq.~\eqref{eq:EulerP}.

The spectral radius is given by
\begin{equation}
r_{sp} = \sqrt{\frac{\gamma * P_{sp}}{\rho_{sp}}} + |u_{sp}|
\end{equation}

Repeating this process at both interfaces yields all required values to solve for a timestep
\begin{equation}
Q_{i}^{n+1} = Q_{i}^{n} + \frac{\Delta t}{\Delta x} (\text{Flux}_{i+1/2}^{n+1/2} -
\text{Flux}_{i-1/2}^{n+1/2})
\end{equation}
The results presented here for the Euler equations use a second-order Runge--Kutta scheme in time, which can be obtained with the same procedure shown in Eqs.~\eqref{eq:RKf} and \eqref{eq:RKm}.

\section*{Acknowledgments}

This material is based upon work supported by NASA under award No.~NNX15AU66A under the technical monitoring of Drs.~Eric Nielsen and Mujeeb Malik.
We also gratefully acknowledge the support of NVIDIA Corporation with the donation of the Tesla K40c GPU used for this research.

\bibliography{GPUPaper}

\begin{thebibliography}{10}
\expandafter\ifx\csname url\endcsname\relax
  \def\url#1{\texttt{#1}}\fi
\expandafter\ifx\csname urlprefix\endcsname\relax\def\urlprefix{URL }\fi
\expandafter\ifx\csname href\endcsname\relax
  \def\href#1#2{#2} \def\path#1{#1}\fi

\bibitem{slotnick:2014}
J.~P. Slotnick, A.~Khodadoust, J.~J. Alonso, D.~L. Darmofal, W.~D. Gropp, E.~A.
  Lurie, D.~J. Mavriplis, {CFD} vision 2030 study: A path to revolutionary
  computational aerosciences, NASA Technical Report, NASA/CR-2014-218178,
  NF1676L-18332 (Mar. 2014).

\bibitem{PattersonLatency}
D.~A. Patterson, Latency lags bandwith, Commun. ACM 47~(10) (2004) 71--75.
\newblock \href {http://dx.doi.org/10.1145/1022594.1022596}
  {\path{doi:10.1145/1022594.1022596}}.

\bibitem{alhubail:16jcp}
M.~Alhubail, Q.~Wang, The swept rule for breaking the latency barrier in time
  advancing {PDEs}, Journal of Computational Physics 307 (2016) 110--121.
\newblock \href {http://dx.doi.org/10.1016/j.jcp.2015.11.026}
  {\path{doi:10.1016/j.jcp.2015.11.026}}.

\bibitem{Alhubail:2016arxiv}
M.~M. Alhubail, Q.~Wang, J.~Williams, The swept rule for breaking the latency
  barrier in time advancing two-dimensional {PDEs}, {\tt
  \href{https://arxiv.org/abs/1602.07558}{arXiv:1602.07558} [cs.NA]} (2016).

\bibitem{BM_BigSolution}
I.~Bermejo-Moreno, J.~Bodart, J.~Larsson, B.~M. Barney, J.~W. Nichols,
  S.~Jones, Solving the compressible {Navier}--{Stokes} equations on up to 1.97
  million cores and 4.1 trillion grid points, in: Proceedings of the
  International Conference on High Performance Computing, Networking, Storage
  and Analysis, SC '13, ACM, New York, NY, USA, 2013, pp. 62:1--62:10.
\newblock \href {http://dx.doi.org/10.1145/2503210.2503265}
  {\path{doi:10.1145/2503210.2503265}}.

\bibitem{Niemeyer:2014hn}
K.~E. Niemeyer, C.~J. Sung, Recent progress and challenges in exploiting
  graphics processors in computational fluid dynamics, Journal of
  Supercomputing 67~(2) (2014) 528--564.
\newblock \href {http://dx.doi.org/10.1007/s11227-013-1015-7}
  {\path{doi:10.1007/s11227-013-1015-7}}.

\bibitem{PascalWP:2016}
{NVIDIA Corporation}, Whitepaper {Nvidia} {Tesla} {P100},
  \url{http://www.nvidia.com/object/pascal-architecture-whitepaper.html}
  (2016).

\bibitem{Witherden20143028}
F.~Witherden, A.~Farrington, P.~Vincent, {PyFR}: An open source framework for
  solving advection--diffusion type problems on streaming architectures using
  the flux reconstruction approach, Computer Physics Communications 185~(11)
  (2014) 3028--3040.
\newblock \href {http://dx.doi.org/10.1016/j.cpc.2014.07.011}
  {\path{doi:10.1016/j.cpc.2014.07.011}}.

\bibitem{MaithamRepo}
M.~Alhubail, Q.~Wang, {KSIDSwept}, git commit \texttt{e575d73},
  \url{https://github.com/hubailmm/K-S_1D_Swept} (2015).

\bibitem{Strzodka}
R.~Strzodka, M.~Shaheen, D.~Pajak, H.-P. Seidel, Cache oblivious parallelograms
  in iterative stencil computations, in: Proceedings of the 24th ACM
  International Conference on Supercomputing, ICS '10, ACM, New York, NY, USA,
  2010, pp. 49--59.
\newblock \href {http://dx.doi.org/10.1145/1810085.1810096}
  {\path{doi:10.1145/1810085.1810096}}.

\bibitem{MalasHager}
T.~Malas, G.~Hager, H.~Ltaief, H.~Stengel, G.~Wellein, D.~Keyes,
  Multicore-optimized wavefront diamond blocking for optimizing stencil
  updates, SIAM Journal on Scientific Computing 37~(4) (2015) C439--C464.
\newblock \href {http://dx.doi.org/10.1137/140991133}
  {\path{doi:10.1137/140991133}}.

\bibitem{VolkovDatta2008}
K.~Datta, M.~Murphy, V.~Volkov, S.~Williams, J.~Carter, L.~Oliker,
  D.~Patterson, J.~Shalf, K.~Yelick,
  \href{http://dl.acm.org/citation.cfm?id=1413370.1413375}{Stencil computation
  optimization and auto-tuning on state-of-the-art multicore architectures},
  in: Proceedings of the 2008 ACM/IEEE Conference on Supercomputing, SC '08,
  IEEE Press, Piscataway, NJ, USA, 2008, pp. 4:1--4:12.
\newline\urlprefix\url{http://dl.acm.org/citation.cfm?id=1413370.1413375}

\bibitem{Gander2015}
M.~J. Gander, 50 years of time parallel time integration, in: T.~Carraro,
  M.~Geiger, S.~Körkel, R.~Rannacher (Eds.), Multiple Shooting and Time Domain
  Decomposition Methods, Vol.~9 of Contributions in Mathematical and
  Computational Sciences, Springer, Cham, 2015, pp. 69--113.
\newblock \href {http://dx.doi.org/10.1007/978-3-319-23321-5_3}
  {\path{doi:10.1007/978-3-319-23321-5_3}}.

\bibitem{falgout2014parallel}
R.~Falgout, S.~Friedhoff, T.~Kolev, S.~MacLachlan, J.~B. Schroder, Parallel
  time integration with multigrid, PAMM 14 (2014) 951--952.
\newblock \href {http://dx.doi.org/10.1002/pamm.201410456}
  {\path{doi:10.1002/pamm.201410456}}.

\bibitem{BABOULIN201217}
M.~Baboulin, S.~Donfack, J.~Dongarra, L.~Grigori, A.~R\'{e}my, S.~Tomov, A
  class of communication-avoiding algorithms for solving general dense linear
  systems on {CPU}/{GPU} parallel machines, Procedia Computer Science 9 (2012)
  17--26.
\newblock \href {http://dx.doi.org/10.1016/j.procs.2012.04.003}
  {\path{doi:10.1016/j.procs.2012.04.003}}.

\bibitem{Anderson}
M.~Anderson, G.~Ballard, J.~Demmel, K.~Keutzer, Communication-avoiding {QR}
  decomposition for {GPUs}, in: Parallel Distributed Processing Symposium
  (IPDPS), 2011 IEEE International, 2011, pp. 48--58.
\newblock \href {http://dx.doi.org/10.1109/IPDPS.2011.15}
  {\path{doi:10.1109/IPDPS.2011.15}}.

\bibitem{Owens:2008ku}
J.~D. Owens, M.~Houston, D.~Luebke, S.~Green, J.~E. Stone, J.~C. Phillips,
  {GPU} computing, Proc. IEEE 96~(5) (2008) 879--899.
\newblock \href {http://dx.doi.org/10.1109/JPROC.2008.917757}
  {\path{doi:10.1109/JPROC.2008.917757}}.

\bibitem{cudaProgGuide}
{NVIDIA Corporation},
  \href{http://docs.nvidia.com/cuda/cuda-c-programming-guide}{{CUDA} {C}
  programming guide}, version 8.0 (2016).
\newline\urlprefix\url{http://docs.nvidia.com/cuda/cuda-c-programming-guide}

\bibitem{Brodtkorb:2013hn}
A.~R. Brodtkorb, T.~R. Hagen, M.~L. S{\ae}tra, Graphics processing unit ({GPU})
  programming strategies and trends in {GPU} computing, J. Parallel Distrib.
  Comput. 73~(1) (2013) 4--13.
\newblock \href {http://dx.doi.org/10.1016/j.jpdc.2012.04.003}
  {\path{doi:10.1016/j.jpdc.2012.04.003}}.

\bibitem{EngineerCuda}
D.~Storti, M.~Yurtoglu, {CUDA} for Engineers: An introduction to
  High-Performance Parallel Computing, Addison-Wesley, 2015.

\bibitem{Cruz:2011gc}
F.~A. Cruz, S.~K. Layton, L.~A. Barba, How to obtain efficient {GPU} kernels:
  An illustration using {FMM} {\&} {FGT} algorithms, Comput. Phys. Comm.
  182~(10) (2011) 2084--2098.
\newblock \href {http://dx.doi.org/10.1016/j.cpc.2011.05.002}
  {\path{doi:10.1016/j.cpc.2011.05.002}}.

\bibitem{FigsShared}
D.~J. Magee, K.~E. Niemeyer, Data, plotting scripts, and figures for
  ``{Accelerating} solutions of {PDEs} with {GPU}-based swept time-space
  decomposition'' (Nov. 2017).
\newblock \href {http://dx.doi.org/10.6084/m9.figshare.4968050.v4}
  {\path{doi:10.6084/m9.figshare.4968050.v4}}.

\bibitem{WangDecomp}
Q.~Wang, Decomposition of stencil update formula into atomic stages, {\tt
  \href{https://arxiv.org/abs/1606.00721}{arXiv:1606.00721} [math.NA]} (2017).

\bibitem{harris:2014}
M.~Harris, {CUDA} pro tip: Do the {Kepler} shuffle,
  \url{https://devblogs.nvidia.com/parallelforall/cuda-pro-tip-kepler-shuffle/},
  accessed: 3 June 2016 (Feb. 2014).

\bibitem{MyRepo}
D.~J. Magee, K.~E. Niemeyer, {Niemeyer-Research-Group/1DSweptCUDA} v2 (May
  2017).
\newblock \href {http://dx.doi.org/10.5281/zenodo.570984}
  {\path{doi:10.5281/zenodo.570984}}.

\bibitem{GlobalLock}
S.~Xiao, W.~C. Feng, Inter-block {GPU} communication via fast barrier
  synchronization, in: 2010 IEEE International Symposium on Parallel
  Distributed Processing (IPDPS), 2010, pp. 1--12.
\newblock \href {http://dx.doi.org/10.1109/IPDPS.2010.5470477}
  {\path{doi:10.1109/IPDPS.2010.5470477}}.

\end{thebibliography}
\bibliographystyle{elsarticle-num}

\end{document}